\documentclass[12pt,preprint]{aastex}

\begin{document}

\title{Observations of Thick Disks in the Hubble Space Telescope Ultra Deep Field}
\author{Bruce G. Elmegreen\altaffilmark{1},
Debra Meloy Elmegreen\altaffilmark{2}\altaffiltext{1}{IBM Research
Division, T.J. Watson Research Center, 1101 Kitchawan Road,
Yorktown Heights, NY 10598; bge@watson.ibm.com}
\altaffiltext{2}{Department of Physics \& Astronomy, Vassar
College, Poughkeepsie, NY 12604; elmegreen@vassar.edu}}

\begin{abstract}
The vertical profiles of chain and spiral galaxies in the Hubble
Space Telescope Ultra Deep Field (UDF) are fit to
sech$^{2}\left(z/z_0\right)$ functions convolved with stellar
profiles in order to measure the disk scale heights $z_0$ in four
passbands. The bulge regions of the spirals are avoided.
Photometric redshifts give absolute scales. The rms heights of the
giant clumps in these galaxies are also measured. The results
indicate that UDF disks are thick with an average $z_0=1.0\pm0.4$
kpc. The ratio of radial exponential scale length to $z_0$ is
$\sim3\pm1.5$. The scale heights are only 20\% larger than the
radii of the giant star-forming clumps and a factor of $\sim10$
larger than the rms clump deviations around the midplanes. This
suggests the clumps formed from midplane gas and dissolved to make
the thick disks. Redshifted stellar population models suggest ages
of $\sim1$ Gy and mass column densities from 4 to 40 M$_\odot$
pc$^{-2}$. The UDF disks look like young versions of modern thick
disks.  This resemblance is difficult to understand if galaxies
grow over time or if subsequent accretion of thin disks
gravitationally shrinks the observed thick disks. More likely,
high redshift disks are thick because their mass column densities
are low; a velocity dispersion of only 14 km s$^{-1}$ reproduces
the observed thickness.  Modern thick disks require more heating
at high redshift. This is possible if the gas that eventually
makes the thin disk is in place before the youngest age of a
modern thick disk, and if the existing stars are heated during the
delivery of this gas.
\end{abstract} 

\section{Introduction}

The Milky Way (Gilmore \& Reid 1983), other spiral galaxies (van
der Kruit \& Searle 1981a,b; Jensen \& Thuan 1982; van Dokkum et
al. 1994; Morrison et al. 1997; Pohlen et al. 2000; Dalcanton \&
Bernstein 2002, see Yoachim \& Dalcanton 2006 and references
therein), and S0 galaxies (Burstein 1979; Tsikoudi 1979; Pohlen,
et al. 2004) have a thick disk component that is old and of
moderate metallicity. In the Milky Way, it has a vertical velocity
dispersion of $\sim35\pm3$ km s$^{-1}$ (Freeman 1987; Chiba \&
Beers 2000) and an exponential scale height between 600 pc (Chen
et al. 2001) and $900\pm300$ (Buser et al. 1999; Ojha 2001; Kerber
et al. 2001; Larsen \& Humphreys 2003). The thin Milky Way disk is
apparently a separate component (Carney et al. 1989; Chen et al.
2001). It varies in height from $\sim60$ pc for the molecular gas
(Cohen \& Thaddeus 1977) to $\sim100$ pc for the young stars to
$330$ pc for the old stars (Chen et al. 2001), with a
corresponding increase in vertical velocity dispersion from
$\sim8$ km s$^{-1}$ for molecular clouds (Stark \& Brand 1989) to
20 km s$^{-1}$ for old stars (Freeman 1987).

These differences in component age, velocity dispersion,
metallicity, and scale height suggest several possible origins for
galactic thick disks (see review in Wyse 2004): (1) they arise
from gradual stochastic heating of stars that form in a thin disk
(Wielen \& Fuchs 1985; Fuchs et al. 2001); (2) they arise from
sudden heating of thin disk stars during the impact of a smaller
galaxy (Quinn et al. 1993; Mihos et al. 1995; Walker et al. 1996;
Robin et al. 1996; Velazquez \& White 1999; Aguerri et al. 2001);
(3) they arise independently of thin disks as tidally stripped
debris from companion galaxies (Gilmore et al. 2002; Abadi et al.
2003; Wyse et al. 2006); (4) they arise as a first step in the
cooling of the disk from a highly turbulent initial condition
(Eggen et al. 1962; Burkert, Truran, \& Hensler 1992; Pardi,
Ferrini, \& Matteucci 1995; Brook et al. 2004, 2005, 2006); and
(5) they arise from the dissolution of massive thin-disk clusters
with small radii and large velocity dispersions (Kroupa 2002).

The distinct and homogeneous nature of thick disks (Seth et al.
2005; Mould 2005) suggests they were well mixed as they formed and
that they stopped forming rather suddenly. This supports models
involving an early phase of satellite and gas accretion, intense
star formation, and vertical mixing that terminated around 6-9 Gy
ago (when $z=0.7-1.4$ in the standard $\Lambda$CDM model). Such
accretion should also give thick disks a significant contribution
from dissolved satellite remnants (Abadi et al. 2003).

The purpose of this paper is to examine the structure of young
thick disks directly, using edge-on galaxies in the Hubble Space
Telescope Ultra Deep Field (UDF).  Reshetnikov, Dettmar, \& Combes
(2003) studied the Hubble Deep Field and noted that high redshift
disks have thicker axial ratios than local galaxies. We found the
same for spirals and chain galaxies in the Tadpole Galaxy
background field (Elmegreen, et al. 2004b, hereafter EEH04) and
the UDF (Elmegreen \& Elmegreen 2005, hereafter EE05; Elmegreen et
al. 2005a; hereafter EERS). Here we measure the disk thicknesses,
colors, and perpendicular color gradients of edge-on UDF galaxies
using distances from photometric redshifts (Coe et al. 2006). We
also measure the perpendicular sizes and rms deviations from the
midplane of the giant star-forming clumps that typically dominate
the structure of young disks. The results are compared to
observations of local galaxies in an attempt to understand the
origin of thick disks. We consider the likely contraction of the
original thick disk if the thin disk accretes at a later time. We
also compare the observations to the predictions of numerical
simulations.

\section{Data}
The Hubble archival UDF images made with 4 filters were used for
this study (observed and processed by S. Beckwith and coworkers in
2004). They include: F435W (B band, hereafter B$_{435}$; 134880 s
exposure), F606W (V band, V$_{606}$; 135320 s), F775W (i band,
i$_{775}$; 347110 s), and F850LP (z band, z$_{850}$; 346620 s).
The images are 10500 x 10500 pixels with a scale of 0.03 arcsec
per px.

The chain and edge-on spiral galaxies studied here were selected
from our UDF catalog (EERS), which included 884 galaxies
classified into 6 morphological types: chain, double, tadpole,
clump cluster, spiral, and elliptical galaxies. There are 114
chain galaxies and 269 spiral galaxies in the catalog. To select
edge-on spirals, we began with the 37 that have an axial ratio
less than 0.3, as determined by ellipticity fits in IRAF. Of
these, 32 were selected for measurement here because they avoid
bright stars and neighboring galaxies. The results from these 32
will be given below. A more refined subsample of 14 edge-on
spirals was considered to include the best cases on the basis of
brightness, size and apparent low inclination; they will be
discussed separately as the ``best-case'' sample. Figure
\ref{spirals6} shows 6 examples of these best spiral cases viewed
in the i$_{775}$ band. For the chain galaxies, we selected 80 good
cases that are relatively isolated in the UDF, and a best-case
subsample of 34 that are bright and well resolved. Figure
\ref{chains6} shows 6 chain galaxies from these best cases at
i$_{775}$ band.

Major and minor axis intensity profiles in each filter were made
using the task {\it pvector} in the Image Reduction and Analysis
Facility (IRAF\footnote{IRAF is distributed by the National
Optical Astronomy Observatories, which are operated by the
Association of Universities for Research in Astronomy, Inc., under
cooperative agreement with the National Science Foundation.}).
This task gives the intensity along a line, averaged over the
width of the line. The major axis profile was several pixels wide
to include the whole galaxy. For the chain galaxies, the minor
axis cut had a width equal to the length of the major axis, again
to include all the light. For the spiral galaxies, two
perpendicular cuts were made, one on each side of the bulge. In
the discussion below, only the average of these two spiral
profiles will be considered, although both were viewed to look for
possible contamination by excessive noise, bright stars, and other
galaxies; when contamination was present, only one side was used.

Scans and Gaussian fits were also made in four passbands for 15
unsaturated stars in the UDF images. The average Gaussian sigma
values were found to be 3.01 px, 3.07 px, 3.02 px, and 3.46 px in
B$_{435}$, V$_{606}$, i$_{775}$, and z$_{850}$ filters,
respectively.  In what follows, these are taken to be the sigma of
the point spread functions, used to deconvolve the instrument
response from the observed galaxy scans.

\section{Analysis}

\subsection{Scale Heights}
\label{sect:sh}

We are interested in the intrinsic widths and lengths of the UDF
galaxies for comparison with the sizes of disk components in
modern galaxies. To measure these quantities in the UDF, we fitted
the perpendicular {\it pvector} scans to convolutions of the star
profiles with certain intrinsic functions, such as exponential and
sech$^{2}$. The sech$^{2}$ function is the equilibrium
distribution of density perpendicular to an infinite isothermal
layer; it is asymptotically exponential at large distance and
Gaussian at small distance. The fit was done by varying the
center, width, and amplitude of the intrinsic function in nested
iteration loops, convolving this function with the star profile,
and then finding the values of the center, width and amplitude
that gave a minimum rms difference between the convolved profile
and the observed scan.

Figure \ref{pvec2625} shows as a solid line (red in electronic
edition) the perpendicular scan for the chain galaxy UDF 7037,
which was illustrated in Figure \ref{chains6}. The star profile is
the narrow (green) line. Two intrinsic functions are illustrated,
an exponential in the top panel and a sech$^2$ in the bottom
panel. The best fit intrinsic functions are shown by dotted curves
and the fits themselves, which are the convolutions of the
intrinsic functions with the star profile, are shown as solid
curves (blue). The vertical scale is linear on the left and
logarithmic on the right. Both exponential and sech$^2$ functions
are good fits in the outer parts (out to the image noise), but the
exponential function is not a good fit in the inner part because
it is too peaked. In what follows, we use the sech$^2$ function to
characterize the perpendicular profiles of the chain and spiral
disks. The resultant scale height is denoted by $z_0$ in the
expression ${\rm
sech}^2(z/z_0)=4/\left(\exp\left[z/z_0\right]+\exp\left[-z/z_0\right]\right)^2$.

The relationship between the sech$^2$ scale height and the
exponential scale height in a fit of vertical light distribution
depends on how far the scan goes into the halo. If the scan
follows the light far into the halo, then the halo region where
the profile is intrinsically exponential dominates the fitted
height and the exponential model gives a value for the height that
equals half the value of the sech$^2$ model. If the scan extends
to only one or two scale heights before it is lost in the noise,
then the two models give more similar values for the height.
Usually galaxy scans include large vertical distances and the
factor of 2 applies; this is the case here as well.  Then the
Milky Way would have a value of sech$^2$ scale height, $z_0$, in
the range from 1.2 kpc to 1.8 kpc (see Sect. I).

Figure \ref{pvep2625} shows the major axis profile of the chain
galaxy UDF 7037, made from a scan 17 pixels wide.  The profile is
not exponential and is dominated by the large clumps visible in
the image in Figure \ref{chains6}. This type of profile is typical
for chain galaxies. Spiral galaxies have smoother major axis
profiles with a central bulge. Other examples of spiral and chain
galaxy profiles are in Elmegreen, et al. (2004a) and Elmegreen et
al. (2005b, hereafter EEVFF).

Figure \ref{pvecrms} illustrates the quality of the scale height
fits for each galaxy. The plus symbols are spirals and the dots
are chains. For each, the large symbols are the best cases and the
small symbols are all of the others in the selected UDF sample.
The abscissa is the perpendicular scale height in pixels and the
ordinate is the ratio of the rms deviation between the fit and the
data divided by the peak value of the intensity scan. These ratios
tend to be slightly smaller for the thicker galaxies, and slightly
smaller also for the best cases, indicating better signal to noise
ratios. Most ratios cluster around 5\%, which we consider to be a
measure of the average fitting accuracy for both spirals and
chains.  The B$_{435}$-band has more galaxies with relative rms
values exceeding 0.1 than the other bands because the galaxies are
often faint in B as a result of their high redshift. The minimum
scale height that can be fitted is $\sim2$ pixels.

The relative error in the fitted scale height was determined by
comparing the best fit height, $z_{0i}$ for galaxy $i$, with the
height that has a relative rms fit different from the best-fit
solution by 10\%, $z^\prime_{0i}$. In both cases the scan centers
and amplitudes were fitted before this height comparison was made.
The relative mean square of the heights around the best fit
heights are given by $\left(1/N\right)\Sigma_{i=1}^N\left(
z_{0i}-z^\prime_{0i}\right)^2/z_{0i}^2$ for summation over $N$
galaxies. The square root of this sum is the relative rms
deviation of the height for a 10\% deviation in the relative rms
of the fit. We considered only galaxies that have relative rms
fits less than 0.1, to avoid drop-out cases and other poor quality
fits. The result indicated that for a relative rms fit deviation
of 0.1, the relative rms height deviation was $0.15\pm0.01$ in all
passbands.  For a relative rms fit deviation of 0.2, the relative
rms height deviation was $0.21\pm0.03$.  Considering the vertical
scatter in Figure \ref{pvecrms}, which is $\sim20$\% of the mean
value of $\sim5$\%, the rms height is considered to be accurate to
this latter deviation of $\sim20$\%.

Figure \ref{pvechhis} shows the distribution function for fitted
scale heights measured in pixels (bottom panels) and kiloparsecs
(top panels) on the z$_{850}$ images. The conversion from pixels
to kpc was done using photometric redshifts in Coe et al. (2006)
and Elmegreen et al. (2006). These redshifts have an uncertainty
of $0.04(1+z)$, but this uncertainty does not affect the kpc
conversion much because the conversion factor is insensitive to
redshift over the redshift range of these galaxies. The samples
are again divided into all of the acceptable cases and the best
cases. There is little difference in the mean results for the
average and best-case galaxies.  The mean z$_{850}$ scale heights
for spiral and chain galaxies are about the same, $4\pm2$ px and
$1.0\pm0.4$ kpc for each. More precisely, the scale heights for
the best case chains and spirals are $0.88\pm0.25$ kpc and
$1.08\pm0.35$ kpc, respectively, and the scale heights for all
chains and spirals are $0.96\pm0.47$ and $1.07\pm0.30$ kpc,
respectively. The mean spiral scale height is $\sim10$\% larger
than the mean chain scale height, but this result is statistically
uncertain.

Figure \ref{pvecred} shows the redshift distributions of scale
height in kpc on the z$_{850}$ images.  Large symbols again denote
the best-case sample. The scale height does not change noticeably
with redshift, although the lowest redshift spiral and the two
lowest redshift chains, both of which are in the best-case
samples, have relatively small scale heights. Perhaps we are
viewing the beginning of the thin disk phase by this late epoch
(or the anomaly could be just a fluke of low number statistics).

\subsection{Color Gradients}

The scale heights in the four passbands were compared to see if
there is a vertical color gradient. Generally the disks were found
to get slightly redder with height. Because the B$_{435}$ image is
often faint and sometimes even blank for the high redshift cases
(the drop-outs), we make this comparison in Figure \ref{pvec2his}
using the V$_{606}$ and z$_{850}$ bands. The plotted quantity is
the $V_{606}-z_{850}$ color fitted to one z$_{850}$-band scale
height off the midplane minus the $V_{606}-z_{850}$ color fitted
to the midplane. For a sech$^2$ function, this color excess comes
from the expression
\begin{equation}
\Delta\left(V_{606}-z_{850}\right)=-5\log\left({{e^1+e^{-1}}\over
{e^{z_{0,Z}/z_{0,V}}+e^{-z_{0,Z}/z_{0,V}}}}\right).
\end{equation}
The figure shows reddening typically between $0.1$ and $0.7$, with
a peak in the distribution for the best-case chains at 0.4 mag,
and a peak in the distribution for the best-case spirals at 0.3
mag. The statistical significance of the 0.1 mag difference
between these two peak color excesses is low. The important point
is that the midplanes are several tenths of a magnitude bluer than
the high latitude emissions.  Because only the ratio of scale
heights for each galaxy determines this result, it is independent
of redshift.

\subsection{Colors and Ages}

Restframe colors were found for as many galaxies as possible by
interpolating magnitudes between the observed passbands and using
the photometric redshifts, which have uncertainties of
approximately $0.04(1+z)$. These redshift uncertainties correspond
to an uncertainty of approximately $\pm0.15$ in
B$_{435}-$V$_{606}$ color, which is about the same as the
photometric uncertainty in the color. We are interested in the
colors at the midplanes and at one scale height from the midplane
using the model fits, so we cannot use NICMOS photometry as the
angular resolution of NICMOS is too poor to resolve the edge-on
disk. Thus we are limited to the galaxies with $z<0.40$ for
restframe B$_{435}-$V$_{606}$ colors. There are 4 galaxies with
such low redshifts, UDF 4907, C2 (from EERS), UDF 3974, and UDF
5454. Their midplane restframe B$_{435}-$V$_{606}$ colors are
$-0.1$, 0.33, 0.71, and 0.35 mag, and their B$_{435}-$V$_{606}$
color gradients from the midplane to one scale height at z$_{850}$
are $-0.16$, 0.28, 0.82, and 0.04 mag, respectively. The errors in
these measurements can be assessed from the ratios of the rms
deviations around the fitted sech$^2$ functions to the observed
peaks (Fig. \ref{pvecrms}). The averages of the ratios for the
B$_{435}$ and V$_{606}$ passbands for these 4 galaxies are 12\%,
8\%, 11\%, and 5\%.  The most accurate values are for galaxies C2
and UDF 5454, which give restframe B$_{435}-$V$_{606}\sim0.34$ in
the midplane. This color corresponds to a population age of
$\sim0.7$ Gy for either an initial burst of star formation or an
exponentially decaying star formation rate with a relatively short
timescale ($<10^8$ yrs). For a decay time of $10^{8.5}$ yrs or
$10^9$ yrs, the age is 1 or 1.9 Gy, respectively. These
conversions from ACS colors to age were done using the spectra in
Bruzual \& Charlot (2003) with the Chabrier (2003) IMF and the
appropriate ACS filters, as described further below.

We also determined B$_{435}-$V$_{606}$ and V$_{606}-$z$_{850}$
colors in the midplane and at one z$_{850}$-band scale height for
all of our galaxies, and compared them with redshifted colors
calculated from population synthesis models. Figure \ref{pvecvzz}
shows the distribution of the observed colors with redshift. The
filled circles and squares are the best cases. The dots and filled
circles are for the midplane values and plus symbols and filled
squares are for the off-plane positions.   The rising color at
high redshift results from redshifting of the Lyman continuum
limit.   There is little vertical color gradient in
B$_{435}-$V$_{606}$ so the symbols overlap in the lower panels.
The V$_{606}-$z$_{850}$ gradient mentioned above is evident in the
displacement between the midplane and off-plane values in the top
panels.

To determine population ages, we reproduce the
B$_{435}-$V$_{606}$, $V_{606}-$i$_{606}$ and i$_{775}-$z$_{850}$
color evolution models from EE05. These models considered a
Chabrier (2003) IMF in the low resolution Bruzual \& Charlot
(2003) spectral tables with $Z=0.008$ (0.4 times the solar
abundance). They included exponentially declining star formation
rates with various decay times starting at $z=6$. To determine the
colors, the population synthesis spectra were redshifted to
certain $z$ values, integrated over time using the star formation
rate as a weighting factor, diminished by absorption from
H$\alpha$ lines (Madau 1995) and dust (Rowan-Robinson 2003;
Calzetti et al. 2000; Leitherer et al. 2002), and then integrated
over the ACS filter functions to get the intensity in each filter
per unit solar mass of stars. The ratios of intensities in
different passbands give the colors.  The results do not differ
significantly for $Z=0.004$.

The models are shown in Figure \ref{pvecvzz} with solid lines for
a duration of star formation equal to 1 Gy and dashed lines for a
duration of 0.5 Gy. The star formation decay times are 0.1 Gy (red
curves in electronic version), 0.3 Gy (green), 1 Gy (blue), and
infinity (continuous star formation, black curves), increasing as
the models get bluer. The midplane data points lie close to the
curves with long decay times, indicating there is still a lot of
active star formation in the midplane. This is true for all
redshifts and colors. The off-plane V$_{606}-$z$_{850}$ colors in
the top panels imply decay times that are shorter than the ages,
0.1 Gy for an age of 0.5 Gy, and 0.3 Gy for an age of 1 Gy.  The
lack of a color gradient in B$_{435}-$V$_{606}$ could be a result
of the closeness of the contours for different decay times.

We conclude that the edge-on disks in the UDF are relatively
young, with dominant stellar luminosity ages in the range from 0.5
Gy to 1 Gy.  There is still significant star formation at all
heights, with slightly more recent star formation in the midplane.
There could be older, fainter stars in these disks as well, so we
cannot get the absolute disk ages from the optical images. At
these redshifts, we see mostly the uv light from young stars in
the galaxy restframes.

For comparison, Yoachim \& Dalcanton (2006) got a mean thick-disk
color of B-R$\sim1.4$ for nearby galaxies. This corresponds to an
age of 10 Gy for the Bruzual \& Charlot (2003) models with either
$Z=0.008$ or 0.004.  Gilmore, Wyse \& Jones (1995) suggested the
Milky Way thick disk is 12 Gy old. These ages are consistent with
the young ages we see in the UDF where the look-back time is 5-12
Gy.

\subsection{Surface Brightnesses}

The distributions of restframe B$_{435}$ and V$_{606}$-band
surface brightnesses at the midplanes of the galaxies are shown in
the bottom panels of Figure \ref{pvecsb0} (solid and dashed lines,
respectively). These come from the sech$^2$ fits to the whole
perpendicular profiles and not the brightness values of the
central pixels (which are blurred by the point spread function and
slightly noisy). There are many more B$_{435}$ values than
V$_{606}$ values plotted because the redshift limits required for
interpolation between the HST bands are $z<0.95$ and $z<0.40$
respectively, and there are few galaxies at low enough redshift to
interpolate into the V$_{606}$ band. The surface brightness levels
are dimmed by cosmological expansion, so the intensities are
corrected in the top panels to rest-frame intensities by
multiplying by $\left(1+z\right)^4$. The restframe corrected
midplane surface brightness for chains ranges from 19.5 to 25 mag
arcsec$^{-2}$ in B$_{435}$ and from 22.5 to 24.5 mag arcsec$^{-2}$
in V$_{606}$. The range for spirals is 21 to 25 mag arcsec$^{-2}$
in B$_{435}$ and $\sim24.75$ mag arcsec$^{-2}$ in V$_{606}$. These
are apparent surface brightnesses, which means edge-on through the
disks of these galaxies. Although the average restframe B$_{435}$
surface brightness is brighter than the average restframe
V$_{606}$ surface brightness for this sample, three of the four
low-redshift galaxies that give restframe V$_{606}$ surface
brightnesses (all chain galaxies) have fainter B$_{606}$
magnitudes, and therefore red colors, as discussed above.

There is a slight correlation between restframe B$_{435}$ surface
brightness and scale height in the z$_{950}$ band, as shown in
Figure \ref{pvecszsb}.  The thicker and generally larger galaxies
have brighter midplane surface brightnesses.  These are edge-on
galaxies, so this correlation is sensible if the midplane surface
brightness is the midplane stellar density multiplied by the path
length through the disk.  The midplane extinction appears to be
lower than in modern disk galaxies, as suggested by the reddening
with height, so the path length through the midplane is larger for
larger galaxies.  We show below that galaxies with larger scale
heights also have larger diameters.

The observed surface brightness in the z$_{850}$ band is shown
versus redshift in Figure \ref{pv2bsbz}.  The dimming with
redshift is slower than the cosmological intensity factor of
$\left(1+z\right)^{-4}$, shown by the curved line. Chains and
spirals have about the same distribution, but the low redshift
spirals are brighter than the low redshift chains. Histograms of
these distributions, compressed along the redshift axis, were
shown in Figure \ref{pvecsb0}.

\subsection{Mass Column Densities}
\label{sect:mass}

We converted the apparent z$_{850}$-band surface brightness (Fig.
\ref{pv2bsbz}) into mass surface density by assuming an age and
star formation decay time of 1 Gy, as estimated above from Figure
\ref{pvecvzz}. We used the ACS filters with the Chabrier (2003)
IMF and the Bruzual \& Charlot (2003) spectral tables for
$Z=0.008$. Galactic dust and intervening hydrogen absorption were
included (see above).

For an apparent z$_{850}$-band surface brightness of 24 mag
arcsec$^{-2}$ and a redshift of 1 (Fig. \ref{pv2bsbz}), the mass
column density measured edge-on through a disk is 24 M$_\odot$
pc$^{-2}$ with 1 Gy age and decay timescales. For a surface
brightness of 25 mag arcsec$^{-2}$ at $z=3$ (also from Fig.
\ref{pv2bsbz}), the mass column density is 125 M$_\odot$ pc$^{-2}$
with these times. With continuous star formation instead of an
exponential decay time of 1 Gy, these column densities become 18
and 78 M$_\odot$ pc$^{-2}$, respectively. For continuous star
formation with an age of 0.5 Gy, they become 12 and  43 M$_\odot$
pc$^{-2}$.

The mass column density perpendicular to the plane may be
estimated from the edge-on value divided by the length-to-width
ratio of $\sim3$, given in Section \ref{sect:lw} below. These
perpendicular values depend on redshift and age assumptions, but
are in a range from 4 to 40 M$_\odot$ pc$^{-2}$. For spiral
galaxies at $z\sim1$, where the average apparent z$_{850}$ surface
brightness ranges between $\sim23.5$ and 25 mag arcsec$^{-2}$ and
the ratio of axes is $\sim2.3$ (see below), the mass column
density for 1 Gy age and 1Gy decay time ranges between 4 and 16
M$_\odot$ pc$^{-2}$.

The surface density of the thick disk in the Milky Way solar
neighborhood may be estimated from the surface density of the old
thin disk ($\sim70$ M$_\odot$ pc$^{-2}$; Freeman 1987), the
density ratio of the thick to the thin disk, which is $\sim5$\%
(Robin et al. 1996; Buser et al. 1998; Ojha 2001; Kerber et al.
2001; Chen et al. 2001; Du et al. 2003) to 13\% (Chen et al. 2001;
Seigel et al. 2002), and the scale height ratio of the thick to
the thin disk, which is $\sim 2$ in these references. The product,
$\sim7-15$ M$_\odot$ pc$^{-2}$, is the thick disk mass column
density perpendicular to the local Milky Way plane. This value is
within the range of values found here for the thick disks of UDF
galaxies.

\subsection{Distribution of Vertical Clump Positions and Sizes in Chain Galaxies}
\label{sect:rms}

\begin{figure}
\epsscale{1.0} \plotone{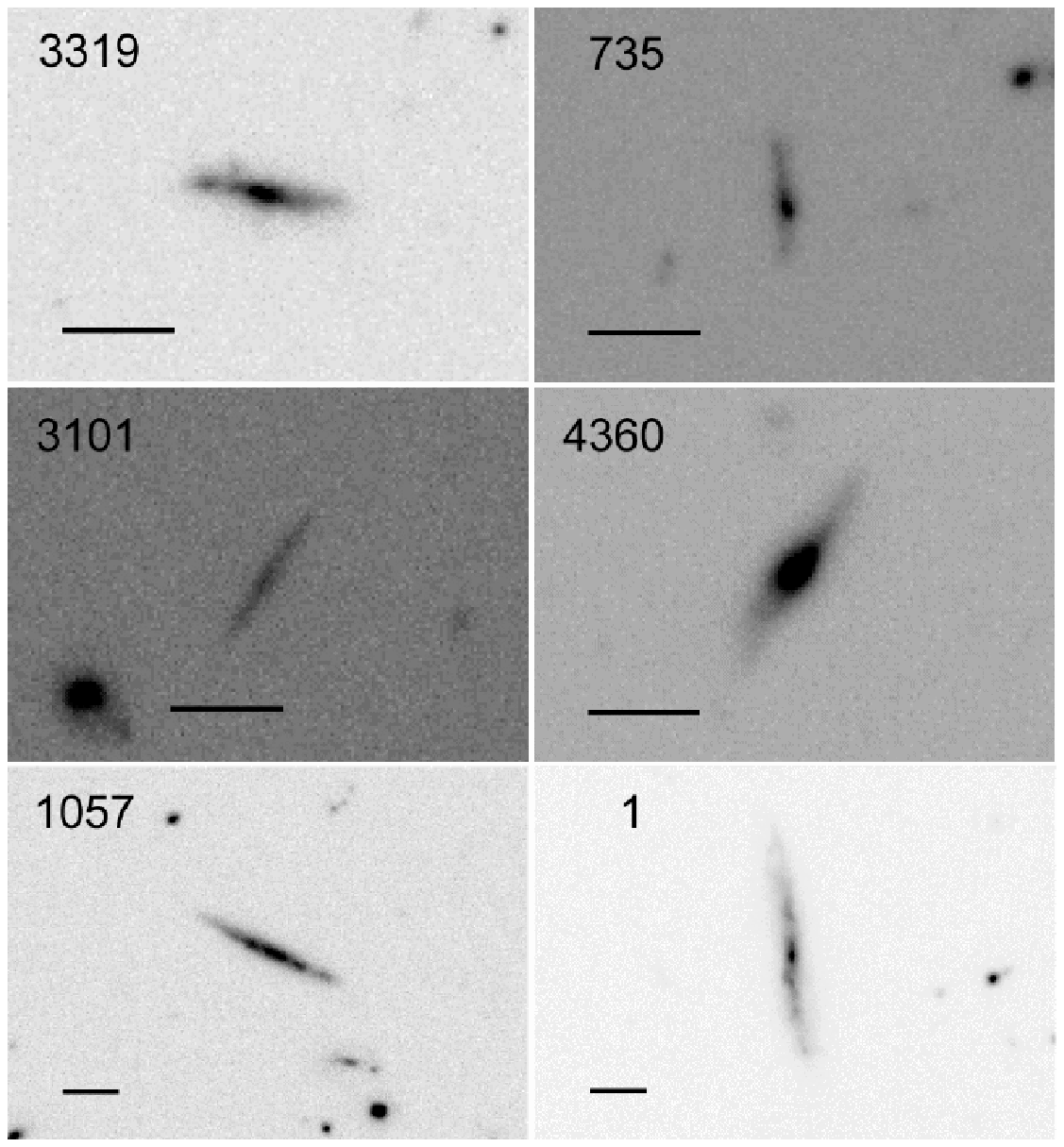} \caption{Sample edge-on spiral
galaxies with UDF catalog numbers indicated. The length bars in
the lower left are 1 arcsec.}\label{spirals6}\end{figure}

\begin{figure}
\epsscale{1.0} \plotone{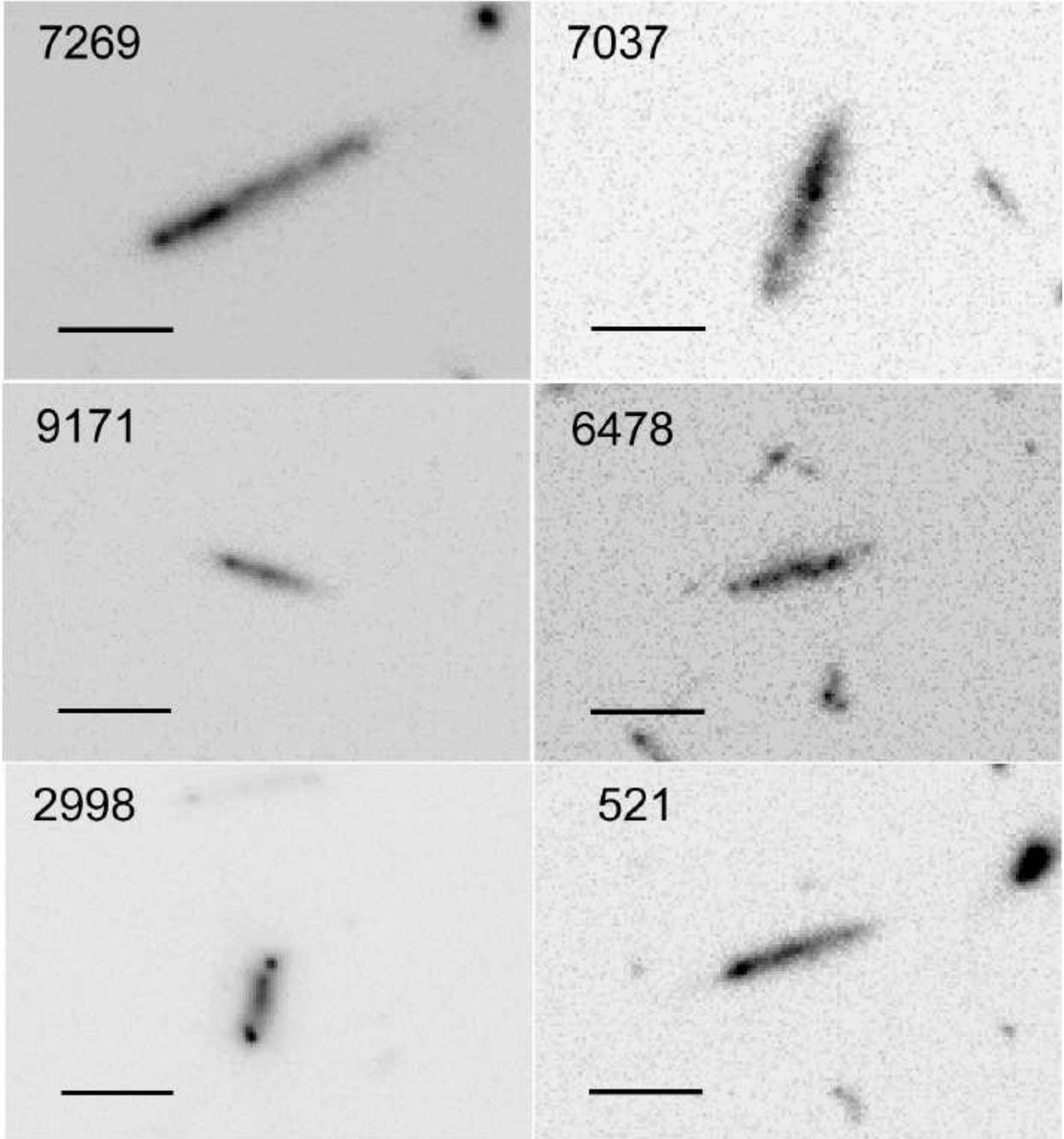} \caption{Sample chain galaxies
with UDF numbers and 1 arcsec scale indicators. Chain galaxies
contain prominent clumps and no central red
bulge.}\label{chains6}\end{figure}

\begin{figure}
\epsscale{0.7} \plotone{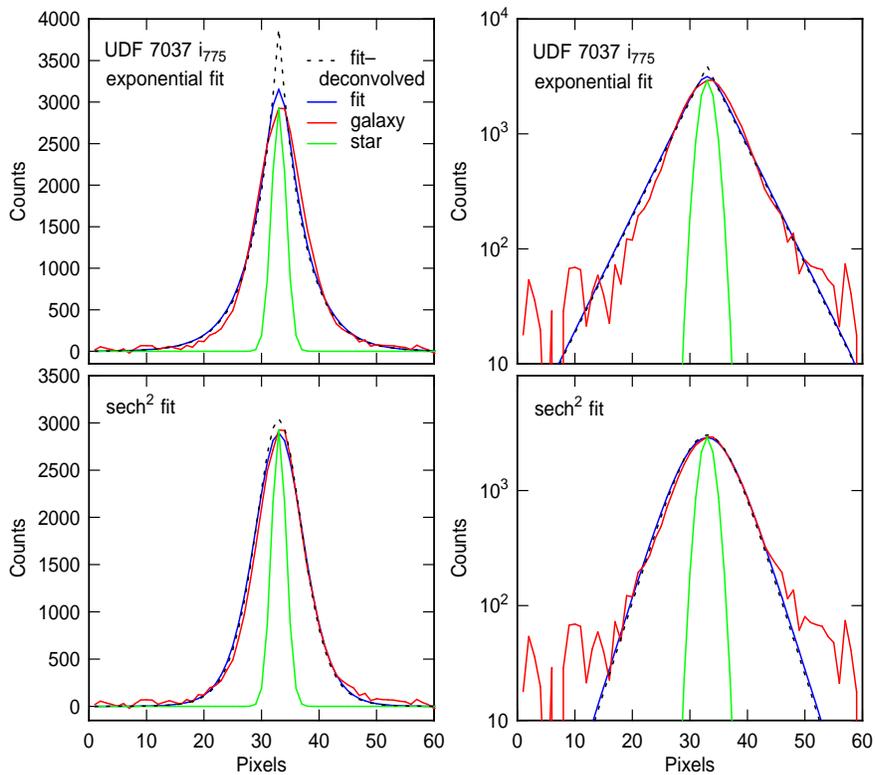} \caption{Perpendicular profiles
through the chain galaxy UDF 7037 in the i$_{775}$ band along with
exponential fits (top) and sech$^2$ fits (bottom). The thin green
profile is the average from 15 stars.  The curve with slight noise
(red) is the galaxy profile, the smooth dashed line (black) is the
intrinsic profile with the star deconvolved (from the fit), and
the solid line (blue) is the convolution of the star and the
intrinsic profile, fitted to the galaxy. Colors are shown in the
electronic version of this manuscript.
}\label{pvec2625}\end{figure}

\begin{figure}
\epsscale{0.7} \plotone{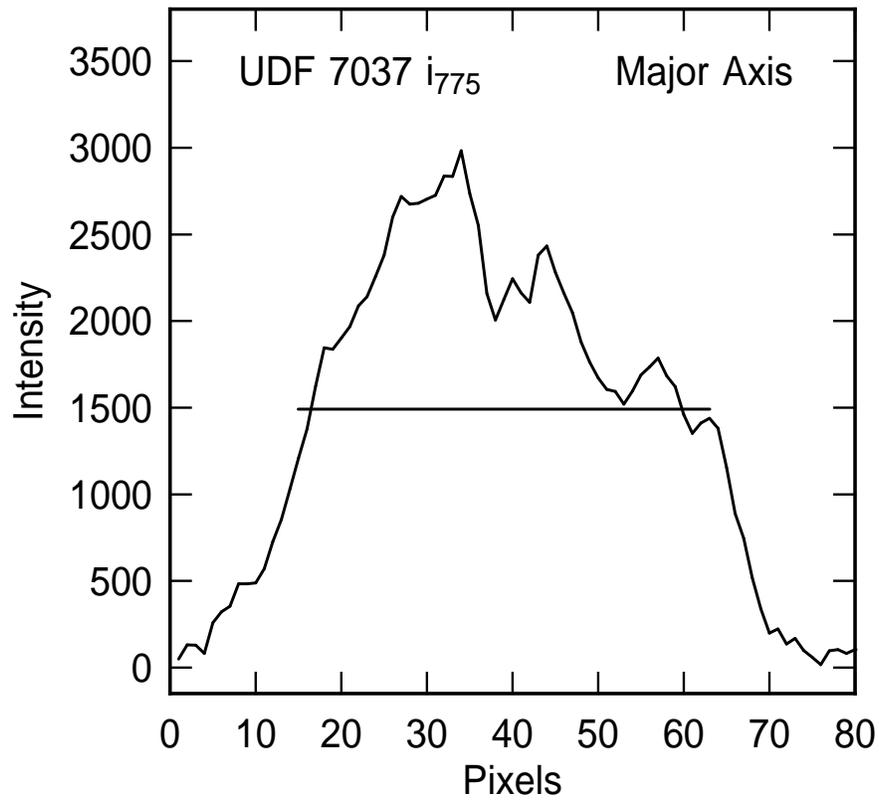} \caption{Parallel profile through
the chain galaxy UDF 7037 in the i$_{775}$ band. The clumps in the
image in  Fig. 2 are the large peaks in the profile. The
horizontal line is the end-to-end length of the galaxy used in
Sect. \ref{sect:lw} and in Figs. 16 and
17.}\label{pvep2625}\end{figure}

\begin{figure}
\epsscale{1.0} \plotone{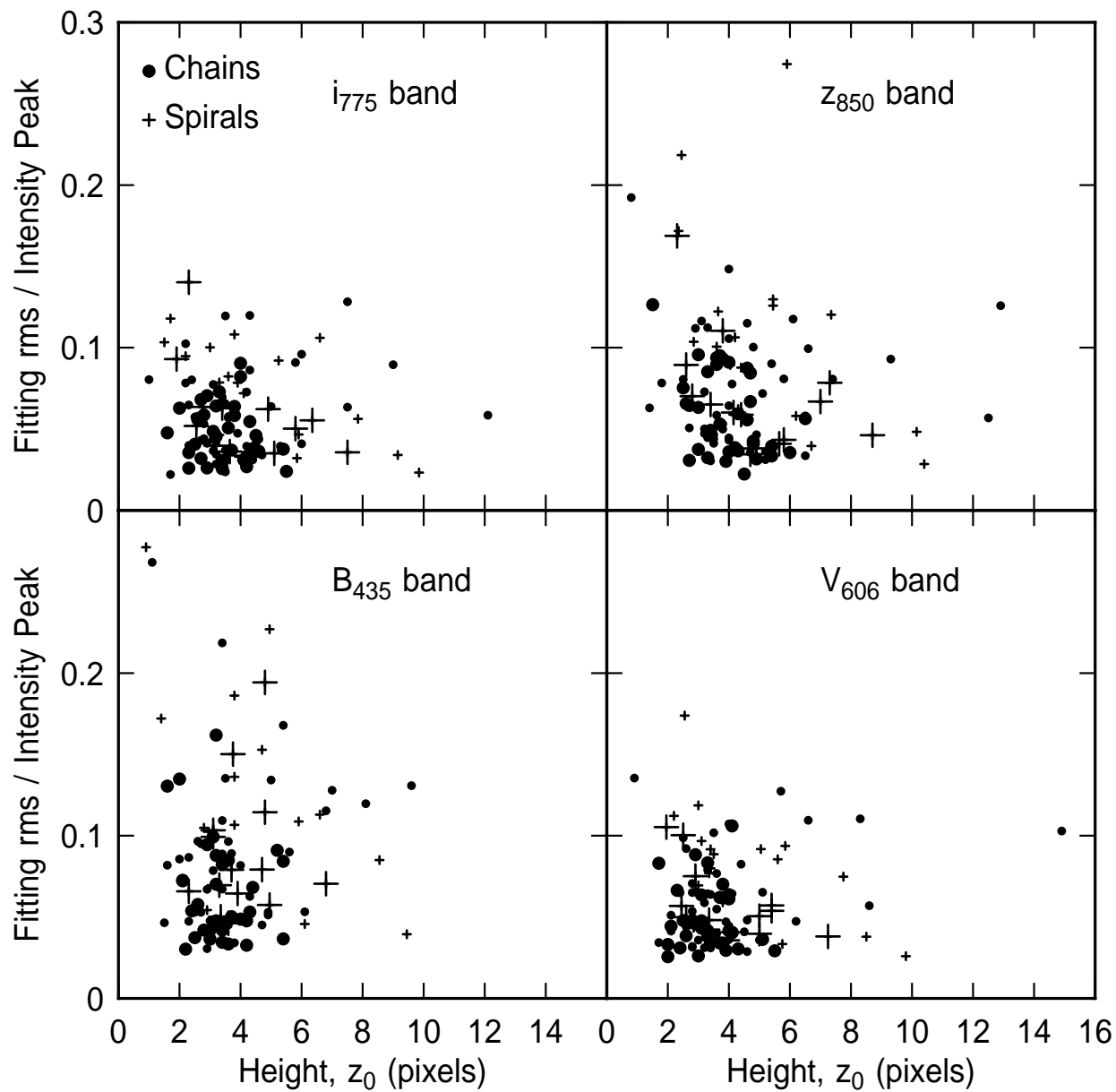} \caption{Ratios of the rms
deviations between observed profiles and the fitted profiles to
the peak intensities in the observed profiles are plotted versus
the derived sech$^2$ heights in pixels. The typical relative
accuracy of the fit is $\sim5$\%. The relative accuracy of the
scale heights is $\sim20$\% (see text). The large symbols are the
best case galaxies.}\label{pvecrms}\end{figure}

\begin{figure}
\epsscale{1.0} \plotone{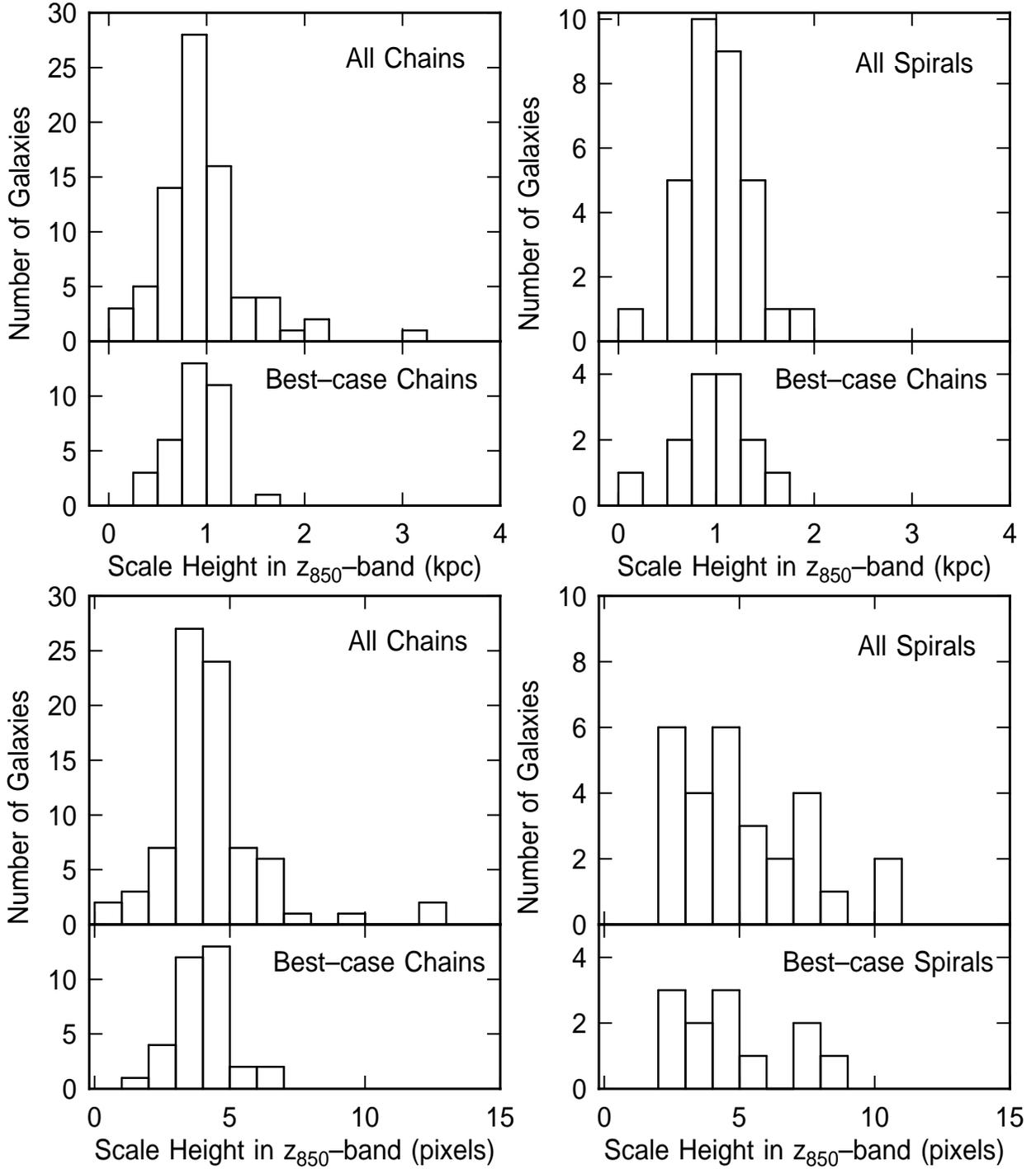} \caption{The distribution of
fitted scale heights, in pixels for the bottom panels and in kpc
for the top panels, assuming photometric redshifts. The
distributions for the best case galaxies are plotted separately.
The chain and spiral galaxy scale heights cluster around 0.9-1
kpc.}\label{pvechhis}\end{figure}

\begin{figure}
\epsscale{1.0} \plotone{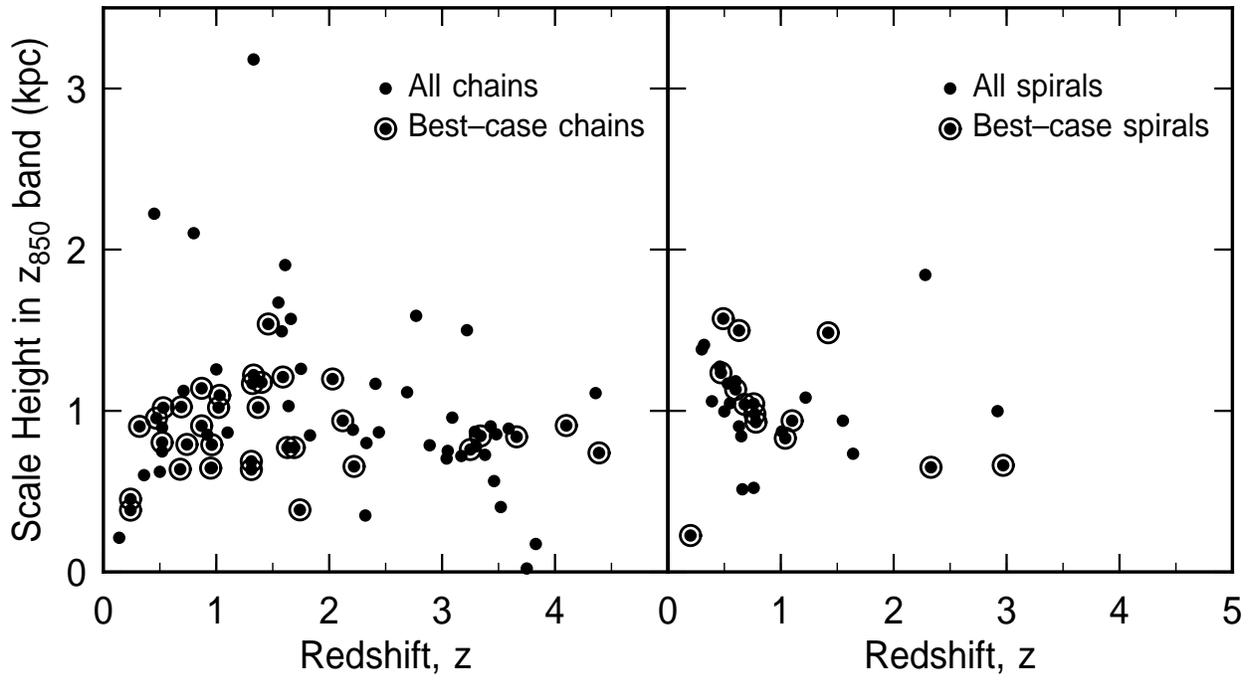} \caption{The z$_{850}$-band scale
height in kpc is shown versus the redshift. There is no systematic
trend except for an excess of small heights at the lowest
redshift.}\label{pvecred}\end{figure}

\begin{figure}
\epsscale{1.0} \plotone{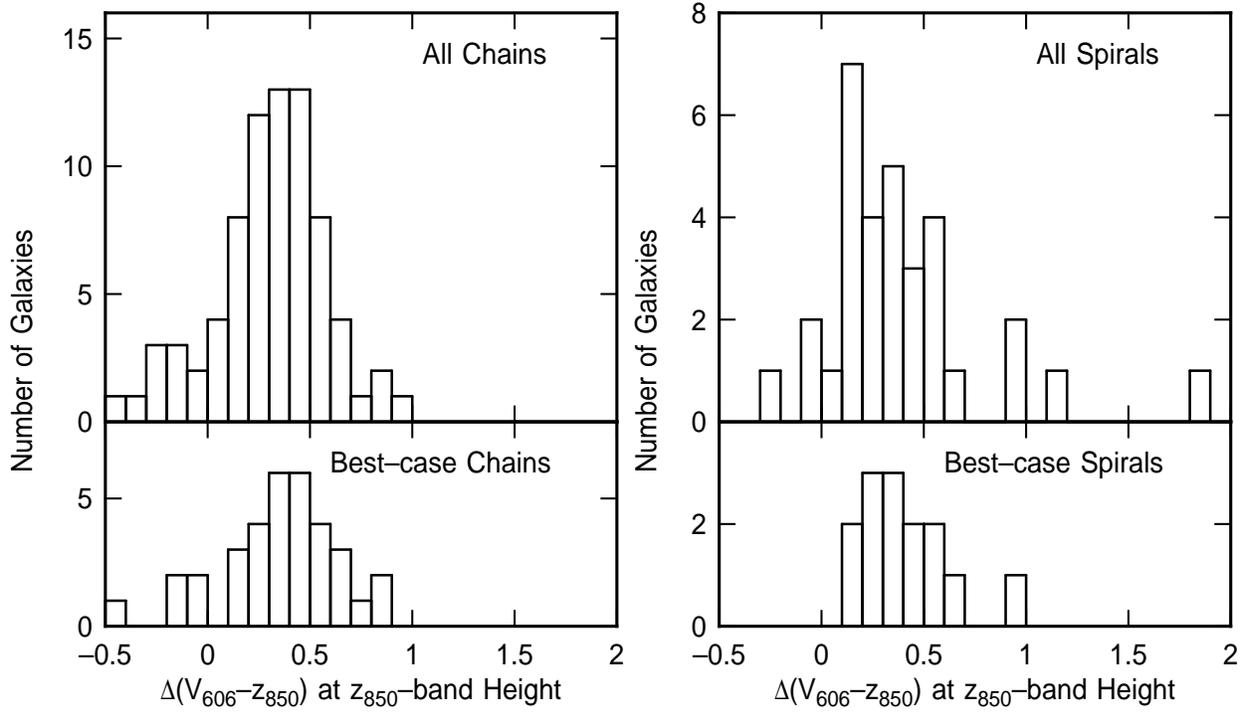} \caption{The distribution of color
difference between the midplane and one z$_{850}$-band scale
height. There is a $\sim0.3-0.4$ mag reddening with
height.}\label{pvec2his}\end{figure}

\begin{figure}
\epsscale{1.0} \plotone{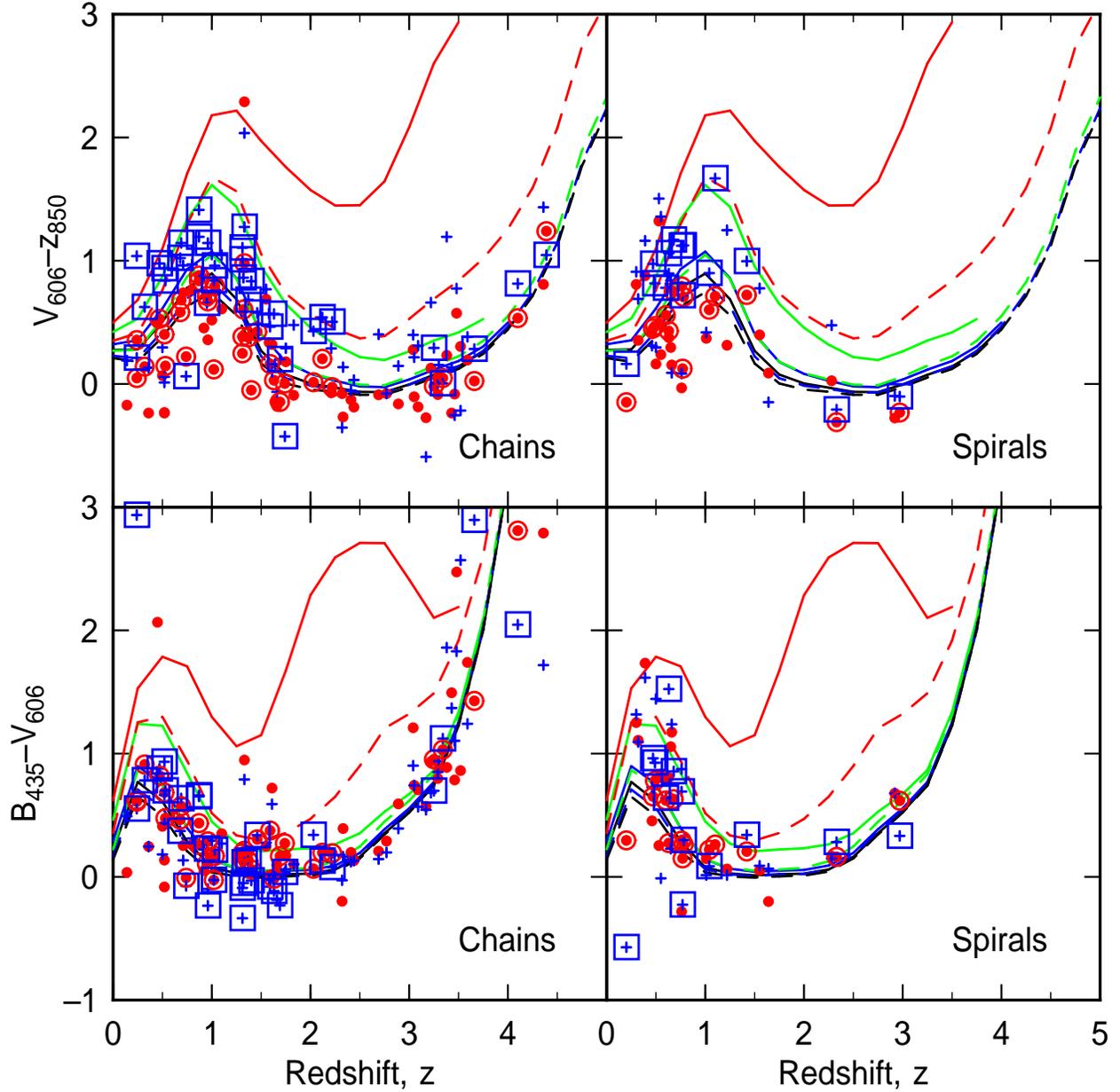} \caption{Color versus redshift
with superposed curves calculated from redshifted stellar
population models. The dots and filled circles (red in electronic
version) are the midplane values, with filled circles for the best
cases. The plus symbols and filled squares (blue) are the
off-plane values, with filled squares the best cases. The solid
lines are for a star formation duration of 1 Gy and the dashed
lines are for a duration of 0.5 Gy. The star formation rate is
assumed to decay exponentially during these times, with decay
times equal to 0.1 Gy (red curves), 0.3 Gy (green), 1 Gy (blue),
and infinity (continuous star formation, black curves), increasing
as the models get bluer (lower in the figure). The best fit curves
suggest there is significant active star formation in all of the
galaxies. The redder off-plane V$_{606}-$z$_{850}$ colors fit to
models with slightly less current star formation. These colors are
consistent with thick disk luminous ages (i.e., weighted toward
these restframe uv bands) of around 1 Gy and either continuous
star formation or slowly declining star
formation.}\label{pvecvzz}\end{figure}

\begin{figure}
\epsscale{1.0} \plotone{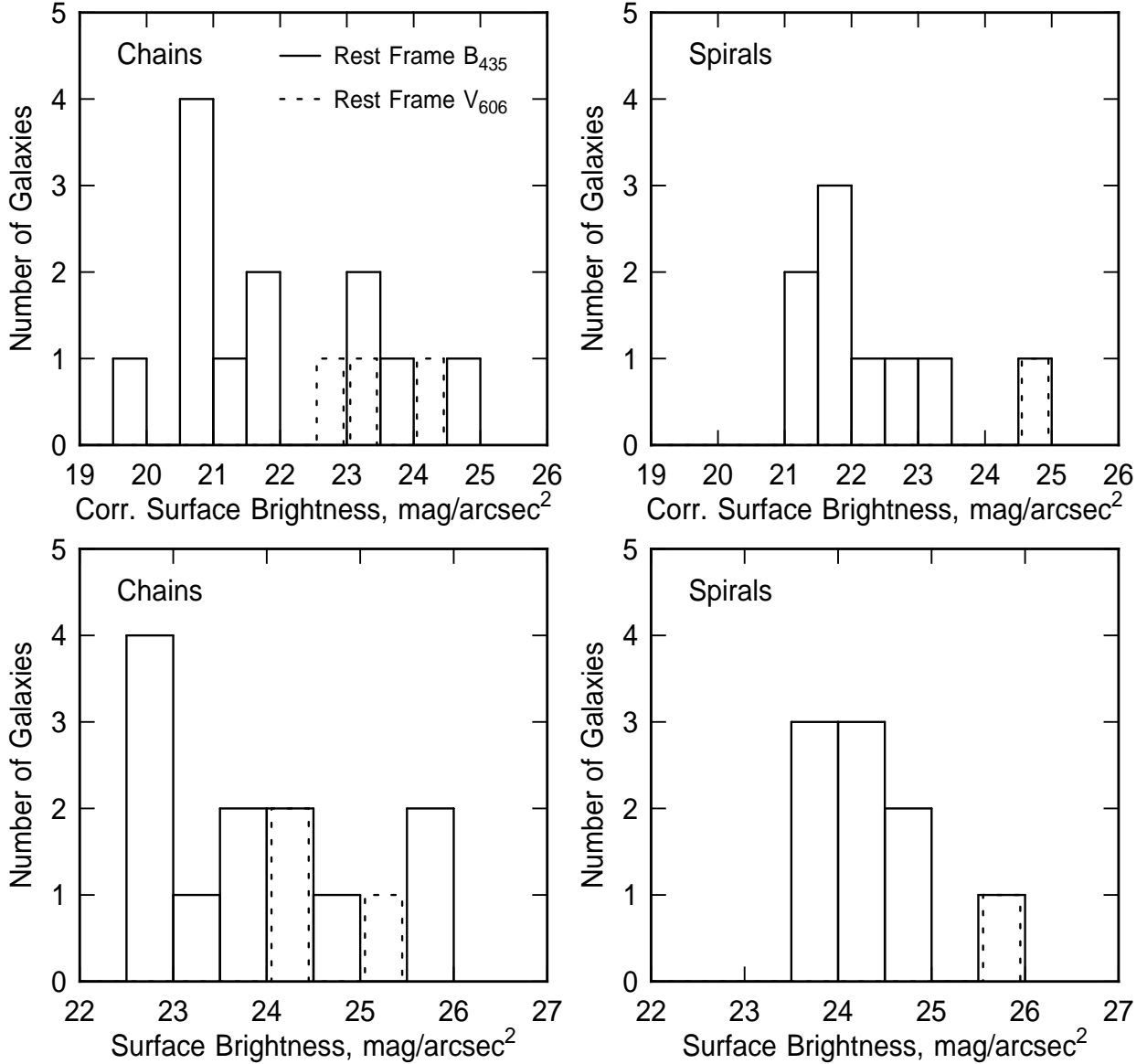} \caption{Observed (bottom) and
corrected restframe (top) surface brightnesses in B$_{435}$ and
V$_{606}$ (dashed lines) at the midplanes of the best case
galaxies. The surface brightness correction is by interpolation
from the ACS bands to the restframe band, and by subtraction of
$10\log\left(1+z\right)$ to account for cosmological dimming. The
midplane values come from the sech$^2$
fits.}\label{pvecsb0}\end{figure}

\begin{figure}
\epsscale{.7} \plotone{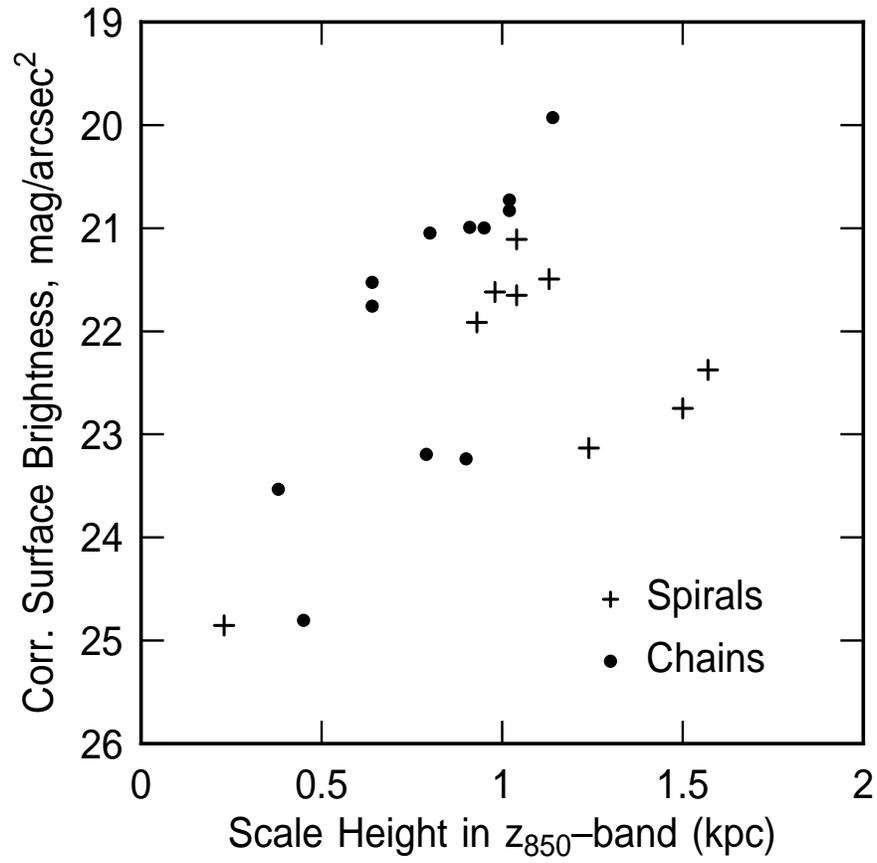} \caption{Corrected restframe
midplane B$_{435}$ surface brightness versus the z$_{850}$-band
scale height. There is a slight correlation in the sense that the
larger galaxies have brighter intrinsic midplane intensities.
}\label{pvecszsb}\end{figure}

\begin{figure}
\epsscale{.7} \plotone{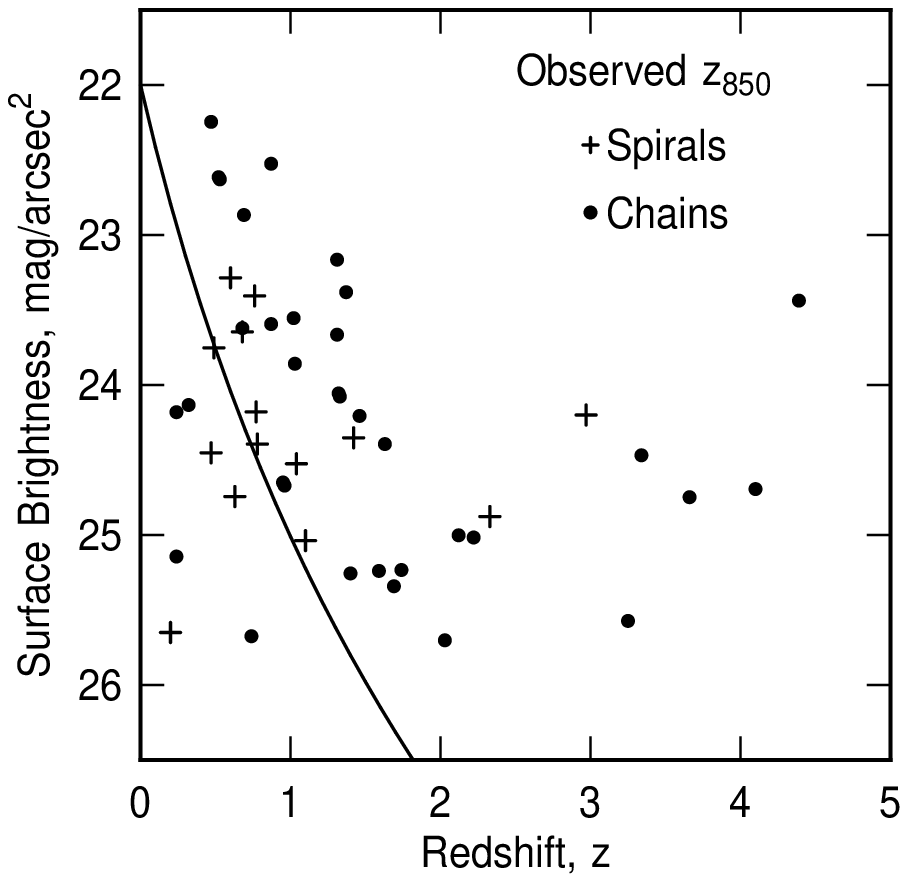} \caption{The fitted midplane
surface brightness in z$_{850}$ band, uncorrected to the restframe
and uncorrected for surface brightness dimming, is shown versus
the redshift. This observed surface brightness is compared to
redshifted stellar population models in the text in order to
estimate the mass surface density of the disk. The curve is the
$\left(1+z\right)^{-4}$ cosmological dimming function with
arbitrary zero point.}\label{pv2bsbz}\end{figure}

Spiral and chain galaxies are very clumpy in the UDF, as indicated
by the images in Figures \ref{spirals6} and \ref{chains6}. The
chains are probably edge-on versions of clump-cluster galaxies,
which were discussed in more detail elsewhere (EEH04; EE05). The
relative emission from giant clumps in all of the UDF clump
cluster galaxies is $0.27\pm0.14$ (EEVFF). For the spirals, it is
$0.08\pm0.065$, not counting the bulges (EEVFF). The spiral clumps
are similar in magnitude and color to the chain and clump-cluster
clumps (EEH04; EE05), so spirals differ by having a more prominent
interclump emission (in addition to bulges and exponential disk
profiles).

We have proposed that galaxy disks evolve along a sequence of
decreasing clump prominence, from clump clusters, chains, and
clumpy spirals at high redshift to normal spirals today (EEVFF).
Some of the clumps in clump cluster and chain galaxies may move to
the nucleus and form a bulge, as suggested by numerical
simulations in Noguchi (1999) and Immeli et al. (2004a,b), but we
see no direct evidence for this and believe it is more likely that
the clumps dissolve where they are. This conclusion comes from the
similarity between internal clump densities and background galaxy
tidal densities, from the appearance in some cases of tails
associated with the clumps, and from the relatively low mass
fraction of the clumps (EE05). The density similarity implies the
clumps are eventually dispersed by in-plane tidal forces.  We also
noted the similarity between the radial distribution of the
average clump luminosity in UDF clump cluster galaxies and the
exponential distribution of UDF spiral disks. All of these results
suggest that the giant clumps in clump cluster and chain galaxies
dissolve more-or-less in place and merge to form a smooth
exponential disk.

\begin{figure}
\epsscale{.9} \plotone{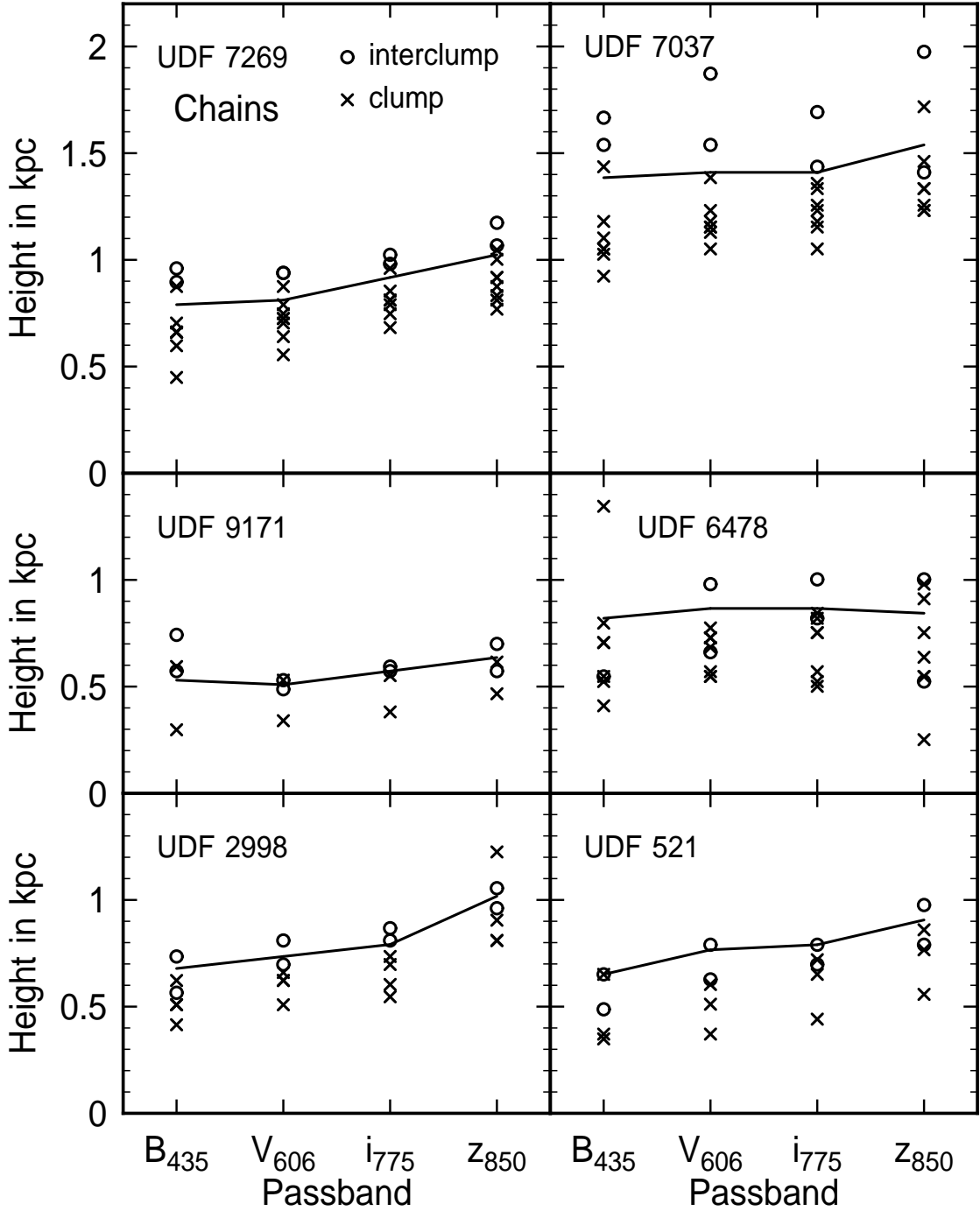} \caption{The sech$^2$ scale
heights of the clumps (x-marks), interclump emission (circles),
and average disks (solid lines) are shown versus bandpass for the
six chain galaxies illustrated in Fig. 2.  The interclump emission
has about the same thickness as the whole galaxy, but the clumps
are $\sim20$\% smaller.  This similarity in size suggests that the
stellar disk is built-up by the dissolution of stellar clumps. The
slight upward trend with longer wavelength is the same reddening
with height shown in Fig. 7.}\label{pvec2cc}\end{figure}

\begin{figure}
\epsscale{1.0} \plotone{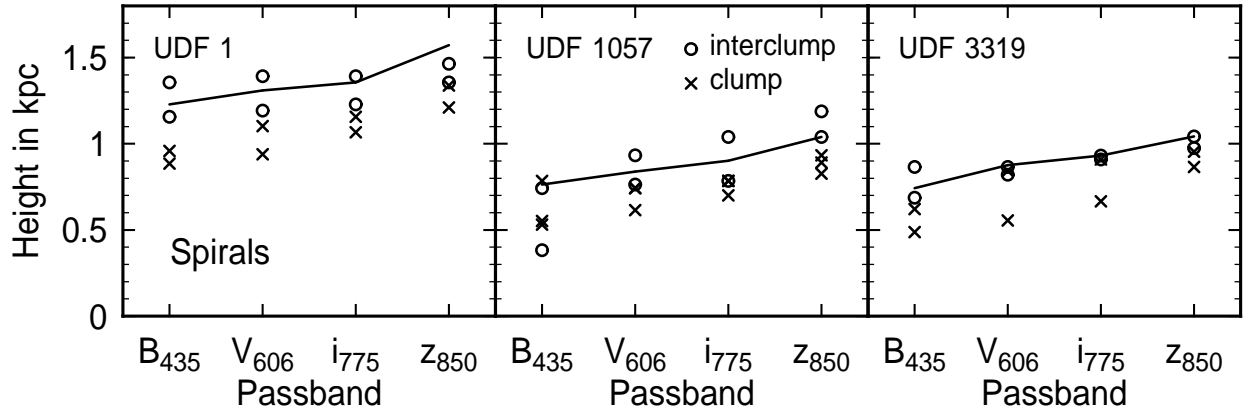} \caption{The sech$^2$ scale
heights of the clumps (x-marks), interclump emission (circles),
and average disks (solid lines) are shown versus bandpass for
three spiral galaxies illustrated in Fig. 1.
}\label{pvec2cs}\end{figure}

The edge-on galaxies in the present study allow us to study the
implications of clump dissolution on the vertical profiles of
galaxy disks.  We did this in two ways. First, the intensity
profiles of all of the prominent clumps in the 6 best-case chain
galaxies in Figure \ref{chains6} were measured in strips 3 pixels
wide.  These scans were also fitted to sech$^2$ functions, using
the same star profile deconvolution technique as for the disks.
The resultant clump scale heights are shown as crosses in Figure
\ref{pvec2cc}, along with the scale heights for the disks, which
are shown as solid lines. The abscissa is the passband and the
ordinate is the height in kpc. For each galaxy and passband, a
different cross is plotted for each different clump.  Interclump
regions were also scanned, and their fitted scale heights are
shown as circles in Figure \ref{pvec2cc}. The heights for clumps
and interclump regions in three of the spiral galaxies of Figure
\ref{spirals6} are shown in Figure \ref{pvec2cs}.

Figures \ref{pvec2cc} and \ref{pvec2cs} show again the tendency
for the disks to redden with height, as the scale heights increase
with wavelength. The same is true for clumps. The figure also
indicates that the clump scale height is $\sim80$\% of the disk
scale height. This near agreement of disk and clump scale heights
is not too surprising considering that the disk profile has a
substantial contribution from the clumps. Nevertheless, the clumps
are about as thick as the whole disks. The clump diameter of 1-2
kpc is only slightly smaller than the in-plane clump diameter,
measured to be 1.8 kpc for a sample of 10 clump cluster galaxies
in the UDF (EE05).  The interclump regions are generally thicker
than the clumps for both chains and spirals, and comparable to the
whole disks.

\begin{figure}
\epsscale{.9} \plotone{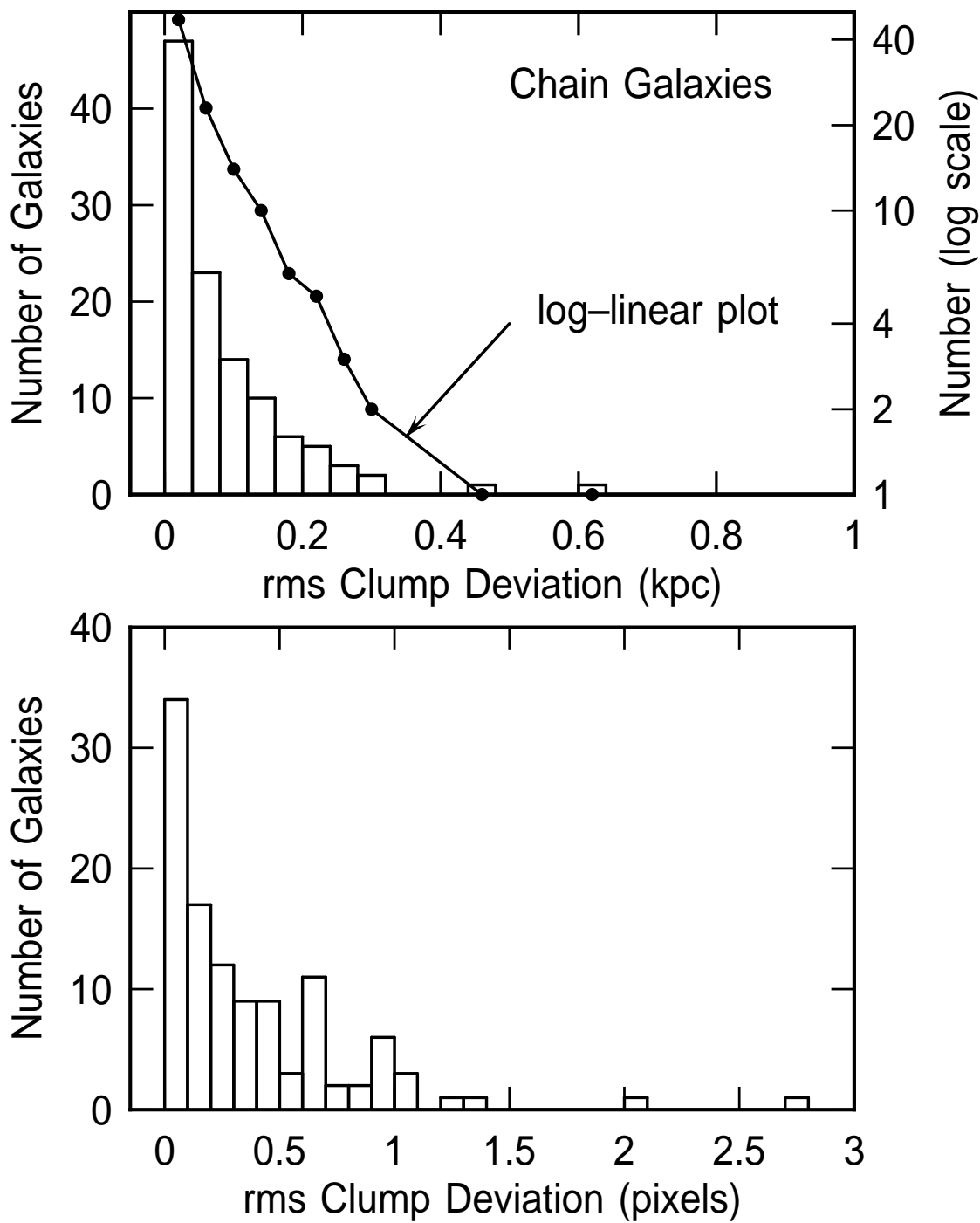} \caption{The distribution of rms
deviation of the clump centers around the average midplane line
defined by these centers, for 431 clumps in 112 chain galaxies
from our UDF catalog. The deviations are shown in pixels at the
bottom and in kpc at the top.  The line at the top uses the
right-hand logarithmic scale. The clump deviations from the
midplane are very small, suggesting the clumps formed there and
did not enter the galaxy from outside.
}\label{pvecclhi}\end{figure}

Secondly, to study vertical clump properties we measured the clump
centroid positions for each chain galaxy in our UDF catalog to one
pixel accuracy. These positions were then fitted to a line for
each chain galaxy (the midplane line) and the perpendicular
distance between each clump centroid and the line was determined.
The rms deviations of these perpendicular distances were
calculated for each chain galaxy. Figure \ref{pvecclhi} shows the
distribution function of the rms deviations in pixels (bottom
panel) and kpc (top panel). Both panels have their histograms
plotted on linear scales, but in the top panel, we also plot the
kpc distribution on a log-linear scale using a solid line; the
corresponding log scale is on the right. The clumps in chain
galaxies typically deviate from the midplane by 0.4 pixel on
average, which corresponds to $\sim100$ pc when the individual
redshifts are applied. This is an upper limit to the physical
deviation because the measuring accuracy is only one pixel. Such a
small deviation compared to the clump scale height is consistent
with our visual impression that chain galaxies are fairly
straight. The solid line at the top of Figure \ref{pvecclhi} is
nearly straight, indicating that the distribution of clump
centroids and their measurement errors combined is approximately
an exponential function of distance from the midplane with a scale
height of $\sim85$ pc.

The confinement of clumps to the midplane seems to imply that they
formed there rather than came in from outside.  If they are
accreted satellites, then they have to settle to the midplane
before they dissolve. In situ formation suggests gravitational
instabilities are involved. We discuss this more in Section
\ref{sect:sum1}.

\subsection{Radial Scale Lengths}
\label{sect:lw}

The parallel {\it pvector} scans for each edge-on spiral galaxy
were fitted to projected exponential functions to find the radial
scale lengths. A projected exponential intensity function $I(x)$
for projected distance $x$ in a disk with unit scale length and
maximum radius $r_{max}$ is
$I(x)=2\int_0^{y^\prime\left(x\right)}e^{-r}dy$ for $r^2=x^2+y^2$
and $y^\prime\left(x\right)=\left(r_{max}^2-x^2\right)^{1/2}$.
Again the fitting procedure iterated over center position, scale
length and peak intensity, finding the minimum rms deviation
between the observed profile and the convolution of the intrinsic
function and the star profile. The radial scale lengths for spiral
galaxies in the UDF were studied in more detail elsewhere (EEVFF).
They are smaller than radial scale lengths in modern galaxies, and
this result, combined with the fact that the UDF spirals have
density waves, suggested that these galaxies grow from the inside
out (EEVFF; see also Papovich et al. 2005). Here we are primarily
interested in the ratio of the radial to the vertical scale
length. The radial scale lengths alone are not as dependable for
edge-on spirals as they were for face-on spirals in our previous
study because for edge-on spirals there is a possibility of
midplane extinction (even though the midplanes are relatively blue
for our galaxies -- see Fig. \ref{pvec2his}).

We also attempted to fit the parallel profiles of chain galaxies
with projected exponential functions, but the intrinsic profiles
are clearly not exponential in most cases (Elmegreen, et al.
2004a; EEVFF) so the resultant scale lengths were sensible for
only a few chains. Instead, we measured the chain galaxy lengths
directly, from the center of the clump at one end to the center of
the clump on the other end.

\begin{figure}
\epsscale{1.0} \plotone{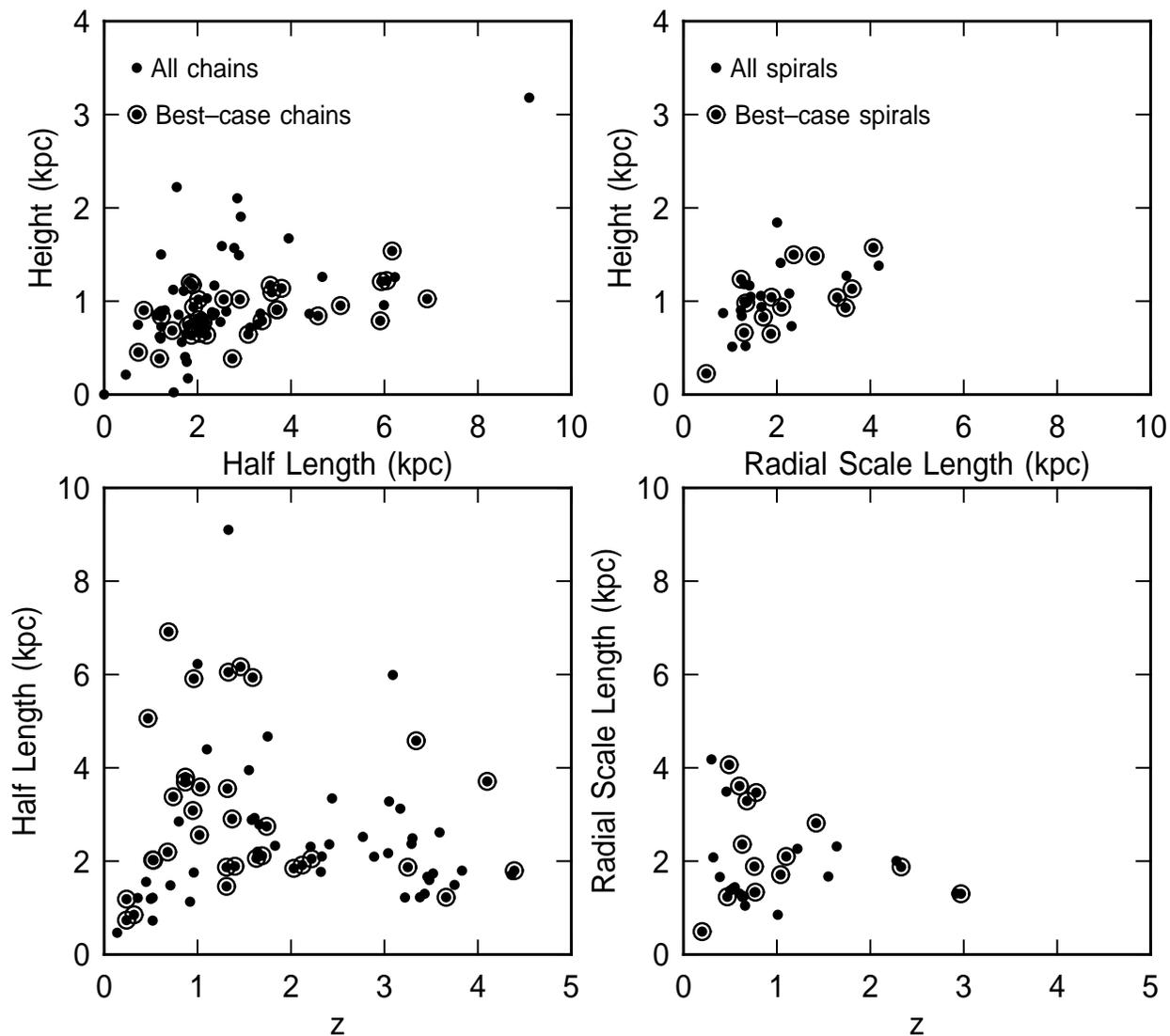} \caption{The radial scale lengths
for spirals (right) and the half-lengths for chains (left), in
kpc, are shown versus the redshift in the lower panels. These are
for the z$_{850}$ band.  The sech$^2$ heights are plotted versus
the radial lengths in the top panels. There is a slight
correlation between the lengths and widths, which makes the ratio
of axis cluster around 3:1 or 2:1 (Fig.
17).}\label{pvecpp}\end{figure}

Figure \ref{pvecpp} shows the radial scale lengths of the spiral
galaxies in the right-hand panel and half the end-to-end lengths
of the chains in the left-hand panel, all for the z$_{850}$ band.
The bottom panels plot these lengths versus the redshift, showing
a slight increase in the largest galaxy size with decreasing
redshift. The top panels plot the vertical scale heights versus
the radial lengths. The scale height increases slightly for larger
galaxies.

\begin{figure} \epsscale{1.0} \plotone{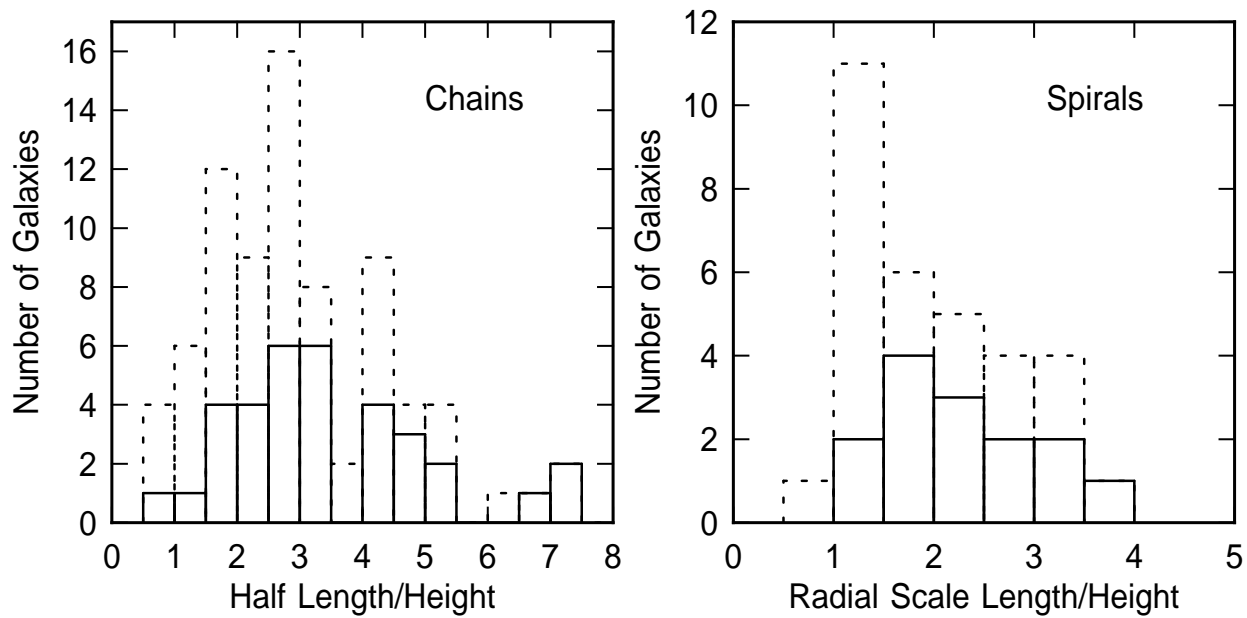} \caption{The distribution
of the ratio of radial scale length (spirals) or half-length
(chains) to perpendicular sech$^2$ height is shown for the
best-case galaxies with solid lines and for all of the galaxies
with dashed lines. The measurements are for the z$_{850}$ band.
Two chain galaxies with less certain measurements are off scale on
the right with ratios of $\sim10$ and $\sim68$.  The average ratio
for the best-case spirals is $2.3\pm0.7$, and the average for the
best-case chains is $3.4\pm1.6$.}\label{pvecpph}\end{figure}

The distribution function of the ratio of the z$_{850}$-band
radial scale length, or half-length in the case of chains, to the
height is shown in Figure \ref{pvecpph}. The solid histograms are
for the best cases, and the dashed histograms are for all the
galaxies. The mean ratio of radial scale length to perpendicular
scale height for the best-case spirals is $2.3\pm0.7$, and the
mean ratio of the half-length to the height for best-case chains
is $3.4\pm1.6$.

These ratios are consistent with our previous results based on the
distributions of axial ratios. For example, the distribution of
width-to-length ratios, $W/L$, was determined for 269 spirals in
the UDF based on outer isophotal contours at 2-$\sigma$ above the
image noise (EERS). This distribution decreases at small $W/L$
with a half-maximum position in the decrease occurring at
$W/L=0.3$ (Fig. 8 in EERS). For local spirals of Types 3 to 8 in
the Third Reference Catalogue of Bright Galaxies (de Vaucouleurs
et al. 1991), this half-maximum point is at $W/L\sim0.1$ (EERS).
The $W/L$ ratio for UDF spirals is slightly smaller than the
inverse of the scale length to scale height ratio for spirals
found in this paper (0.4) but here we corrected for projection
effects. If we fit the observed intensity profile along the major
axes of our spirals by an exponential function, rather than a
projected exponential function, then the average ratio of scale
length to scale height would be $3.1\pm1.1$. The inverse of this
is more in line with the $W/L$ ratio, which also has no projection
corrections. For chains and clump-cluster galaxies combined (i.e.,
different orientations of the same objects), the distribution of
$W/L$ decreases at small $W/L$ with a half-maximum position
occurring at $W/L\sim0.2$ (EE05). This $W/L$ based on 2-$\sigma$
isophotal contours is a slightly different measure than our ratio
of half-length to scale height here, but the two are consistent
with each other, and both indicate relatively thick disks for
chain and clump-cluster galaxies.

\section{Discussion}

\subsection{Summary and Implications of the Results}
\label{sect:sum1}

The edge-on galaxies in the UDF have disks that are $1.1\pm0.4$
kpc thick for spirals and $0.9\pm0.3$ kpc for chains.  No thin
disk components or dust lanes can be seen, although resolution
limits would make this difficult (a pixel is typically $\sim250$
pc). There is probably some thin disk component in the spirals
because there are spiral density waves in the face-on versions;
such waves propagate best in a disk that is thinner than the
interarm spacing. There are no spirals in clump cluster galaxies,
however, so if chain galaxies are the edge-on counterparts of
clump-clusters, then it is possible the chains do not have thin
disk components inside the observed thick disks.  This difference
between spirals and chains is consistent with our proposal that
chain and clump-cluster galaxies are young versions of spiral
galaxies (if they survive major mergers; EEVFF).

The restframe and observed colors of the galaxies suggest active
star formation in the midplane and slightly less active star
formation off the midplane. Color ages are in the range from 0.5
Gy to 1 Gy, with star formation rates either continuous during
that time or slowly decaying. For these ages, the edge-on surface
brightnesses correspond to perpendicular mass column densities
between 4 and 40 M$_\odot$ pc$^{-2}$ depending on redshift and
galaxy size; for spirals it is between 4 and 16 M$_\odot$
pc$^{-2}$. The column density of the Milky Way thick disk in the
solar neighborhood is comparable, $\sim7-15$ M$_\odot$ pc$^{-2}$.

There is a gradual reddening of the UDF disks with height, by
0.3-0.4 mag in V$_{606}-$z$_{850}$ over one scale height, which is
consistent with the lack of thin and prominent dust layers. This
gradient corresponds to $0.3-0.4$ mag kpc$^{-1}$ in restframe U-B
for $z\sim1$. Local thick disks have smaller or no reddening
gradients with height. For example, Mould (2005) found a vertical
gradient in V-I of $\sim0.06\pm0.02$ mag kpc$^{-1}$ for four local
galaxies.  Our high redshift color gradients should be larger than
local color gradients for several reasons. First, the restframe
U-B colors seen at high redshift by the ACS are very sensitive to
stellar population age for ages less than $\sim1$ Gy, but the
restframe V-I colors seen locally are less sensitive to $\pm1$ Gy
formation times when the population has already aged 10 Gy. For a
Bruzual \& Charlot (2003) model with a Chabrier (2003) IMF and
metallicity $Z=0.008$, U-B increases by 1.0 mag as a starburst
ages from 0 to $5\times10^8$ years.  However, the increase in V-I
between 9.5 Gy and 10 Gy old populations, which have the same
$5\times10^8$ yr star formation age spread, is only 0.11 mag.
Thus, young U-B color gradients in ACS galaxies should evolve into
much smaller V-I color gradients in today's galaxies. A second
change over time is that the stars in thick disks at high redshift
should mix vertically during the intervening minor mergers.
Similarly, the thick disks in major merger remnants today should
be a mixture of the thick disk and other stars from the previous
galaxies.

There is no obvious trend in disk thickness with redshift, and
only a slight trend in disk length with redshift. The ratio of
radial scale length to disk height is $2.3\pm0.7$ for spirals.
Exponential fits along radius could not be made for the chains, so
we measured instead the ratio of the half-length to the disk
height, which is $3.4\pm1.6$. These measurements are consistent
with other observations of pervasive thick disks at high redshift.
The implication is that most or all galaxy disks were thick at
early times, from $z=0.5-4.5$ in our sample, which corresponds to
look-back times of 4.8 Gy to 12.2 Gy in the standard cosmology.

Such pervasive disk thickness is not surprising given the small
mass column densities in the disks we observe (Sect.
\ref{sect:mass}) and in the original Milky Way thick disk. Using
the equilibrium relation for sech$^2$ scale height
$z_0=a^2/\left(\pi G\Sigma\right)$, we expect the observed
$z_0\sim1$ kpc for a velocity dispersion of only $a=14$ km
s$^{-1}$ when the mass column density is comparable to that of the
Milky Way thick disk, $\Sigma\sim15$ M$_\odot$ pc$^{-2}$.  In
comparison, the velocity dispersion of the Milky Way thick disk is
much larger, $\sim35$ km s$^{-1}$, because of the additional disk
mass from the thin disk. These considerations suggests that most
high reshift disks should be thick for star formation at normal
interstellar velocity dispersions, and that local thick disks may
have a qualitatively different origin (i.e., they require much
more heating). We discuss this comparison more in the next
section.

The UDF disks have giant clumps that are probably star-forming
regions with ages of $\sim100-300$ My (EE05). These clumps are
centered within $\pm100$ pc of the midplane for our complete
sample of chain galaxies. This small dimension is partly a
selection effect because chains are defined to be linear objects.
There are other clumpy UDF galaxies with more irregular shapes
that could have intrinsically larger rms clump positions. In any
case, the chains as a class appear to have rather relaxed vertical
structures, and the sizes of the clumps that comprise them are
only $\sim20$\% less than the disk scale heights.  This relative
flatness and the similarity in clump and disk sizes suggest that
the clumps form inside the disks and then dissolve in place to
form the disks. This is analogous to star formation in modern thin
disks but the star cluster scale in the UDF ($0.4-1$ kpc in radius
according to Fig. \ref{pvec2cc}) is much bigger than the star
cluster scale in today's galaxies.  A more relevant comparison
might be between modern star complexes and UDF clumps, because
both are at the top of a hierarchy of young stellar structures.
Star complexes in local galaxies measure $\sim400-800$ pc along
the plane and $\sim100$ pc perpendicular to the plane (Efremov
1995). They are often separated by $\sim2$ kpc along the arms of
large spiral galaxies.  These in-plane dimensions are comparable
to the in-plane dimensions of UDF clumps, but the vertical
dimensions are $\sim10$ times larger in the UDF. Star complexes
most likely form by gravitational instabilities in today's thin
disks, with a separation and characteristic in-plane scale given
by the ambient Jeans length for a disk, $\lambda\sim2a^2/G\Sigma$.
With velocity dispersion $a\sim10$ km s$^{-1}$ and gas mass column
density $\Sigma\sim20$ M$_\odot$ pc$^{-2}$, this gives
$\lambda=2.3$ kpc for modern disks.  The same in-plane separation
should apply to high redshift disks but it should also apply
approximately to the vertical direction at high redshift because
the whole disk column density at high redshift is small, not just
the gas disk column density as in modern galaxies.

\subsection{More Comparisons with Modern Galaxies and a Consideration of
Adiabatic Thick Disk Contraction} \label{sect:comp}

The thick disks at high redshift should also be compared to modern
thick disks in their absolute heights and length-to-width ratios.
In the largest recent survey, Yoachim \& Dalcanton (2006)
determined the scale heights for thick disk components of local
galaxies using the same sech$^2$ function as that used here and
found an average relation $z_0=1.4\left(V/100\;{\rm km
\;s}^{-1}\right)$ for circular rotation speed $V$.  These modern
values are larger than ours by a factor of $\sim1.5$ on average,
but the modern galaxies have larger radial sizes too. The ratio of
radial exponential scale length to vertical sech$^2$ scale height
in the Yoachim \& Dalcanton sample, $3.4\pm1.7$, is about the same
as ours, which is $2.3\pm0.7$ for spirals and $3.4\pm1.6$ for
chains. Thus the UDF scale heights are about the same as local
scale heights for similarly small galaxies. This size similarity,
along with the observed young ages, suggests we are witnessing the
formation of the thick disk component in young galaxies. The
confinement of the star-forming clumps to the midplanes and their
similarity in size to the scale heights of the thick disks
suggests further that the mechanism of thick disk formation is the
dissolution of giant star clusters. These clusters may form by
gravitational instabilities, as illustrated by  Noguchi (1999) and
Immeli et al. (2004a,b) under similar conditions.

There are three problems with this simple interpretation of the
observations. First, the galaxies are smaller than local galaxies,
on average (EEVFF), and if they are supposed to grow by accretion
into modern L$^*$ galaxies, then both their radial scale lengths
and thicknesses have to grow in order to keep the ratio of axes
about the same.  If the Yoachim \& Dalcanton (2006) scaling is
applied to the Milky Way where $V\sim220$ km s$^{-1}$, the
vertical scale height would be $z_0\sim3$ kpc, meaning it has to
grow by a factor of 3 from the observed 1 kpc average at high
redshift. The radial scale length has to grow as well, by a factor
of $\sim4.4$ if the $\sim2.3\pm0.7$ ratio of axes for high
redshift spirals converts to a $3.4\pm1.7$ ratio for modern
spirals (Yoachim \& Dalcanton 2006). This mutual growth implies
that if the thick disk formation process continues to be through
the dissolution of massive clusters, there should be stars in
modern thick disks that are fairly young, going back only to the
time when the radial growth stopped, if it ever did. As thick
disks appear to be mostly old today (older than 10 Gy for the
Milky Way -- Gilmore, Wyse, \& Jones 1995; Furhmann 1998), they
could not have grown for as long a time as the thin disk.

Second, a high fraction of the galaxies we observe in the UDF
could merge together or into larger galaxies over time (see merger
trees in Brook et al. 2005). Any such merging would destroy the
structures we observe here. However, merging would not cool a hot
disk, so the thick disk stars in the present sample are likely to
end up as thick disk stars in the merger remnant, or perhaps as
halo stars if the merger is violent. After the last major merger,
the thick disk component of the final galaxy should contain the
previous thick disk components of the earlier galaxies. Such thick
disks would not be simply the superposition of dissolved giant
star clusters, and modern thick disks that form in this way would
not be expected to resemble the high redshift thick disks that we
observe here (i.e., with midplane clumps having the same size).

\begin{figure} \epsscale{.7} \plotone{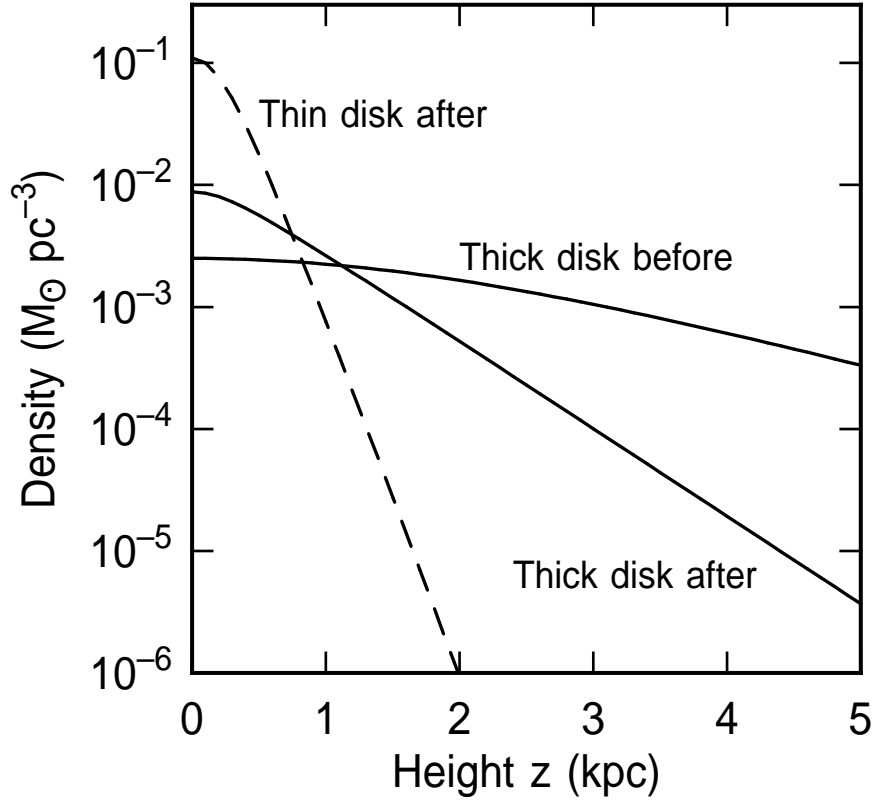}
\caption{Model illustrating the vertical contraction of an initial
thick disk after the addition of a massive thin disk component.
The thick disk has a constant mass column density of 15 M$_\odot$
pc$^{-2}$ but its velocity dispersion increases adiabatically
during the contraction. Before the addition of the thin disk, the
thick disk has a sech$^2$ scale height of 3 kpc and a vertical
velocity dispersion of 25 km s$^{-1}$. The thin disk column
density is 70 M$_\odot$ pc$^{-2}$ and its velocity dispersion is
18.5 km s$^{-1}$.  The initial thick disk shrinks down to a
smaller thick disk component with a $1/e$ half-thickness of 875
pc, a vertical velocity dispersion of 38 km s$^{-1}$, and a
midplane mass ratio to the thin disk of 8\%. This final state
resembles the old thin disk and the thick disk of the Milky Way,
suggesting that the initial thick disk had to be $\sim3$ kpc in
scale height -- much larger than what we observe in the
UDF.}\label{pvecmod}\end{figure}

Third, any slow accumulation of young thin disk material inside a
thick disk will pull in the thick disk because of the extra
midplane gravity.  Figure \ref{pvecmod} shows a solution to the
two-fluid coupled equations:
\begin{equation}
dP_i/dz=-g\rho_i \;\;\;;\;\;\;P_i=\rho_i
a_i^2\propto\rho_i^\gamma\;\;\;;\;\;\; g(z) =4\pi G\int_0^z\left(
\rho_0+\rho_1\right)dz\label{dpdz}
\end{equation}
where $i=0$ and 1 for thin and thick disk material. The pressure
equilibrium equations are separate but the gravitational forces of
the two fluids are coupled (see also Jog 2005).  The figure also
shows the single fluid solution representing an initial pure-thick
disk that has the same mass column density as the ``thick disk
after'' solution. Both the single fluid solution (``thick disk
before'') and the thin$+$thick disk combined solution (``after'')
are equilibria.

To make this solution, we varied the initial thick disk scale
height and the final thin disk velocity dispersion ($a$) until the
final thick$+$thin disk resembled the Milky Way. We assumed that
the thick disk column density remains constant at 15 M$_\odot$
pc$^{-2}$ (see above) as the colder thin disk material is added.
We assumed the final thin disk column density is 70 M$_\odot$
pc$^{-2}$, as in the Solar neighborhood, and the ratio of midplane
densities for thick and thin disks in the final state is 8\%,
which is also typical for measurements locally.  The thick disk
velocity dispersion is allowed to increase adiabatically during
its contraction with the usual three dimensional adiabatic index,
$\gamma=5/3$ (this assumes that stellar heating during the
one-dimensional contraction is distributed in all three
dimensions).  The velocity dispersion of the cold component is
assumed to be constant, and fitted to 18.5 km s$^{-1}$ to give the
desired final state.  The other variable was the initial thick
disk scale height, from which the initial velocity dispersion
follows given the assumed thick disk mass column density.  This
initial height was found to be 3 kpc for the best final-state fit,
giving an initial velocity dispersion of 25 km s$^{-1}$ for the
assumed mass column density.

These solutions suggest that an initial thick disk of 15 M$_\odot$
pc$^{-2}$, with a sech$^2$ scale height of 3 kpc and a 25 km
s$^{-1}$ velocity dispersion, can contract to a final disk
component representing 8\% of the midplane density with a $1/e$
height (not sech$^2$ height) of 875 pc and a velocity dispersion
of 38 km s$^{-1}$ when a thin disk component is slowly added,
where the thin disk has a final $1/e$ height of 350 pc and a
velocity dispersion of 18.5 km s$^{-2}$. These numbers for the
final state resemble the thin and thick disk of the local Milky
Way.  The problem is that the initial pure-thick disk in the model
has a sech$^2$ height of 3 kpc, which is much larger than we
observe in the UDF (where the average is 1.0 kpc).

Gnedin et al. (2004) considered a similar problem of halo
contraction during disk build-up for a Milky Way model. They found
that the density of the halo increases by a factor of $\sim5$
within $\sim1$ kpc of the galactic center and that this increase
is comparable to that obtained in a simple adiabatic model. Figure
\ref{pvecmod} has a similar density increase in the central part
of the thick disk.

To turn the problem around, we considered what would happen to an
observed 1 kpc thick disk with a mass column density of 15
M$_\odot$ pc$^{-2}$ when a 70 M$_\odot$ pc$^{-2}$, cold ($\sim20$
km s$^{-2}$) thin disk accretes through it.  The new equilibrium
would have the original thick disk contract to a scale height that
is the same as the final thin disk scale height, $\sim350$ pc, and
the resulting velocity dispersion for both would be $\sim20$ km
s$^{-2}$. They would appear as a single component with a small
fraction of very old stars in addition to the young thin-disk
stars (old thin disk stars were also found in simulations by Abadi
et al. 2003 for a different reason). Unless the thin disk mass is
already present inside the UDF galaxies, the thick disks we are
seeing in the UDF are probably not the precursors of today's thick
disks, even though they look about the same as today's thick
disks.  They could be precursors of the old thin disk component of
today's disks, mixed in with whatever material accreted in the
intervening time.

The model shown in Figure \ref{pvecmod} suggests that if today's
thick disk component of the Milky Way started as an equilibrium
pure-thick disk at a young age, and if subsequent disk accretion
was gentle (adiabatic), then the initial thick disk scale height
had to be $\sim3$ kpc, much larger than we observe in the UDF. The
surface brightness of this component would be lower than what we
see by the ratio of heights (for a fixed mass column density),
which is $1/3=0.3$. This corresponds to a lower surface brightness
by 1.3 mag.  The surface brightness can be even fainter if the
original thick disk was older than the disks observed here.
Perhaps such ultra-thick components are present and we cannot see
them. They would have an axial ratio of unity so they are more
like spheroids, not disks. It is possible that early disk galaxies
had both thick disks and spheroids $\sim3$ times higher, and that
accretion over a Hubble time brought in the original spheroid
components to become modern thick disks, as it brought in the
original thick disks to become the modern old thin disks, blending
them with the accreted gas and stars over time. This all assumes
adiabatic gas accretion, however, and no disk stirring by
satellites that might bring in the gas. It also assumes today's
halo stars came in relatively late on satellites.

An alternative model is that the galaxies we observe in the UDF
are not going to accrete or grow much over time, but will remain
small galaxies with the observed disks becoming today's thick disk
components. There would have to be relatively little thin disk
accretion. This solution is plausible for these small UDF galaxies
because modern galaxies of this size, the dwarf Irregulars and
spheroidals, tend to have thin disk components with low surface
densities (see Fig. 22 in Yoachim \& Dalcanton 2006).  Then we
would conclude that the UDF does not contain enough galaxies to
sample out far enough in the galaxy luminosity function to produce
an $L^*$ galaxy, but is mostly composed of common dwarfs.  A
possible problem with this interpretation is that the spirals in
the UDF have strong spiral arms, whereas dwarf galaxies today do
not. This implies that the ratio of visible disk mass to halo mass
is larger in the UDF spirals than in today's dwarfs, and makes it
unlikely that the UDF spirals will turn into today's small
galaxies (EELS).

These considerations of shrinking disks do not apply if the thick
disk components of modern galaxies have been heated steadily over
time, either by satellite stirring or satellite debris deposition.
What would change in equations \ref{dpdz} is the scaling of the
velocity dispersion with density, which was assumed to be
adiabatic. Any disk thickness can result if there is enough
stirring. A problem with this continuous stirring model arises if
today's thick disks are mostly old stars. Stirring without any
deposition or triggering of young, high-velocity stars would seem
to be difficult. If the stirring stopped at the time given by the
ages of the youngest thick-disk stars, and subsequent accretion
was adiabatic thereafter, then the shrinking argument applies
again for subsequent times.  This would seem to be the case in the
Milky Way models reviewed by Freeman \& Bland-Hawthorn (2002), who
suggest that the Milky Way thin disk grew quiescently after
$z\sim1$. In that case, the youngest thick-disk stars would be
$\sim8$ Gy old.  Quiescent growth by a significant mass factor
would still shrink the original thick disk though.

Smooth disk accretion is favored by the observed confinement of
the giant clumps, particularly in the chain galaxies, to the
galaxy midplanes (Sect. \ref{sect:rms}).  In this case the clumps
most likely formed by gravitational instabilities in the disk gas
(as opposed to whole clump accretion from outside).  Smooth gas
accretion is considered in galaxy formation models by Murali et
al. (2002), Westera et al. (2002), Sommer-Larsen, G\"otz, \&
Portinari (2003), and Keres et al. (2004).

In the Introduction, we reviewed the main models for the formation
of thick disks. There are elements of all of these models in our
observations of the UDF and in the analogous observations of
modern thick disks: (1) Stochastic heating is likely for young
stars at low velocity dispersion, particularly with giant
star-forming clumps like those in the UDF as scattering centers.
(2) Sudden heating during impacts from small galaxies is not
observed directly in the present sample of galaxies, but it is
likely to have happened at $z\ge1$, as collisions were probably
frequent. (3) Satellite debris is not likely to be the primary
origin for the thick disks in the UDF because they appear too much
like dissolved giant clumps. However, if the UDF thick disks
shrink down during a later phase of thin disk accretion, then a
new, late-stage formation mechanism would be required to form
modern thick disks.  This later mechanism could be the
accumulation of satellite debris, particularly as the remnant
streams are apparently observed in the Milky Way (e.g., Wyse et
al. 2006) and there should have been many residual satellites.  A
counter-rotating thick disk supports this possibility too (Yoachim
\& Dalcanton 2005).

(4) Thick disk formation during a highly turbulent phase, before
the accreted material cools to form a thin disk, is essentially
what we observe in the UDF. High turbulent speeds can produce the
observed giant clumps by gravitational instabilities. We note,
however, that if the thick disk column density is comparable to
that in the Milky Way and typical for our observations, namely
$\Sigma\sim15$ M$_\odot$ pc$^{-2}$, then a velocity dispersion of
only $a=14$ km s$^{-1}$ is enough to produce a thick disk with the
observed UDF sech$^2$ height of $z_0=1$ kpc (from the equilibrium
equation $a^2=\pi G\Sigma z_0$). Young disks should always be
relatively thick before a significant column accumulates. In
addition, the impact heating of disk gas and high accretion
velocities for new gas can maintain some excess turbulence over
that produced internally by supernovae and stellar winds. When the
accretion rate and star formation rate decrease and this excess
turbulence subsides, and when the disk mass increases, then the
following generations of young stars should form a thinner disk.

Finally, (5) the thick disks in the UDF appear most directly to
have formed by the dissolution of massive star clusters. The
difference between this interpretation and the model by Kroupa
(2002) is that the star clusters in the UDF have very large radii,
rather than large velocity dispersions with radii that are small,
like those of today's clusters.

We also note that the earliest interpretation of the modern thick
disks as inner compressed parts of old stellar halos (van der
Kruit \& Searle 1981a,b), could be possible too, as discussed
above.

Several questions are raised by the UDF observations. Is there a
component of the old thin disk in the Milky Way and other spiral
galaxies that is the compressed remnant of a thick disk that was
in place at $z\ge1$?  Is there a range of ages in the thick disks
of modern galaxies that suggests they grew for some extended time
along with the thin disk as part of an inside-out formation
scenario?  Are the ages of thick disk stars younger in the outer
parts of galaxies (which formed last)? Was the accretion onto
disks at early times mostly in the form of gas (rather than stars
or stellar satellites) so that young stars formed {\it in situ} by
the usual processes (i.e., midplane gravitational instabilities)?
Could the modern thick disk component have been a yet-undetected
spheroidal component at high redshift, compressed into a flatter
form by the subsequent accretion of thin disk mass? All of these
questions are within the realm of modern telescopes and numerical
simulations.

\subsection{Comparisons with Numerical Simulations of Thick Disk
Formation}

Disk galaxy simulations show many of the features reported here,
but there some notable differences too. Here we review several of
the recent thick disk models.

Abadi et al. (2003) found in simulations of Milky Way type
galaxies that the thick disks are composed mostly ($>50$\%) of
stars brought in by satellites; among the oldest thick-disk stars
($>10$ Gy), 90\% are from satellites. They also found that the
thin disk forms more quiescently after the last significant merger
is over by converting the previously accreted satellite gas into
stars. These simulations are consistent with our general results,
but they do not discuss the attributes of the disks at early
epochs, when we observe them here. Thus, we cannot directly
compare our results to theirs.

Models by Brook et al. (2004) form thick disks from gas-rich
building blocks that are accreted during hierarchical merging. The
gas forms stars while it cools into a disk. This differs from the
Abadi et al. model in which most of the thick disk stars formed in
satellites before they accreted. Brook et al. (2004) showed images
of the model during the epoch of thick disk formation $\sim8$ Gy
ago. The models look very clumpy, like clump cluster galaxies, but
the clumps do not settle into a thick disk until rather late, 8.3
Gy ago ($z\sim1.2$) for the model they show. We suspect that real
clump cluster galaxies are more disk-like even at higher redshift,
based on the distribution of the ratio of axes (EERS) and the
continuation of this morphology out to $z\sim5$ (Elmegreen et al.
2006). We also find here that the rms dispersion of the clump
positions around the midplanes of the chain galaxies are
immeasurably small, unlike the model in Brook et al. (2004). Thus
our conclusion about the origin of these clumps, i.e. that they
arise from in-plane instabilities, differs from the model in Brook
et al, where they are directly accreted from outside. If clumpy
accretion models eventually show more disk settling before the
clumps dissolve, then an accretion origin might be able to explain
the highly aligned morphology of the chain galaxy clumps. Brook et
al. (2004) also find, as in Abadi et al., that the epoch of the
last major merger determines the age of thick disk.

Brook et al. (2005) consider the thick disk component of their
models in more detail. They see no vertical color gradient after
the thick disk has evolved to the present time. As mentioned in
Section \ref{sect:sum1}, this color uniformity is consistent with
the strong vertical gradient that we do see in the restframe U-B
color when the thick disks are young because evolution and
bandshifting remove any such gradient over time. However, the
Brook et al. result is also consistent with no vertical gradient
for a young age because of vertical mixing. Thus the lack of a
vertical color gradient in the present-day Brook et al. models is
inconclusive. It would be interesting to know the model color
gradients at $z\sim1-5$ where the chain galaxies in our survey are
located. There should be a vertical gradient in the restframe U-B
of $\sim0.4$ mag kpc$^{-1}$. Such gradients would seem to be a
critical test of whether thick disks at high redshift formed by
mixing satellite debris (where small early gradients are expected)
or by quiescent star formation at low disk surface density (where
larger gradients might appear; Sect. \ref{sect:sum1}).

The Brook et al. (2005) models have a relatively high abundance of
alpha elements in the thick disk because of rapid star formation.
We observe thick disk colors consistent with relatively young ages
too, $\sim0.5-1$ Gy (Fig. \ref{pvecvzz}).

Further implications of the Brook et al. models are in Brook et
al. (2006), who study a Milky Way-like disk after redshift $z=1$.
They find that the exponential scale length increases as redshift
decreases from $z=1$ to 0, by about a factor of 1.4, and that the
vertical scale height does not change during this time. We have
essentially the same result here but over a longer redshift range.
We find that the z$_{850}$-band scale height is somewhat constant
from $z=4$ to 0 while the scale length or half-length (in the case
of chains) increases a little. However the scatter in our data is
too high to be conclusive about this. More to the point, though,
is the change in the ratio of the length to the height for spirals
in our survey. This ratio is $2.3\pm0.7$ from Figure
\ref{pvecpph}, and $3.4\pm1.7$ for modern galaxies in Yoachim \&
Dalcanton (2006). The change is a factor of 1.5, similar to the
prediction by Brook et al. (2006).   There is essentially no
change in this ratio for chain galaxies, but the scatter in the
data is large.

Brook et al. (2006) also find that the central surface brightness
of their model disk in B band at $z=1$ is 21.0 mag arcsec$^{-2}$.
The dimming-corrected restframe B-band surface brightness found
here is fainter, 22 mag arcsec$^{-2}$ (Fig. \ref{pvecsb0}) for the
edge-on spirals, and about 1 mag arcsec$^{-2}$ fainter if we
correct to a face-on view. The corrected restframe B-band surface
brightness for chain galaxies is a little brighter, but still not
as bright as in the Brook et al. (2006) model.

\section{Summary}

The transverse thicknesses of chain galaxies and edge-on spirals
in the UDF were fitted to sech$^2$ functions and the resulting
scale heights were studied as a function of redshift and galaxy
size. The average scale height is $1.0\pm0.4$ kpc, with no obvious
dependence on redshift. There is a $0.3-0.4$ mag
V$_{606}-$z$_{850}$ color gradient over one scale height,
suggesting slightly older stars at higher elevations. The midplane
colors corrected for redshift or compared with redshifted stellar
population models indicate star formation is still active.
Luminous disk ages are $0.5-1$ Gy, although older stars could be
present. The redshift-corrected surface brightnesses, compared
with population models, suggest the mass column density
perpendicular to the disk is in the range from 4 to 40 M$_\odot$
pc$^{-2}$, depending on galaxy size and redshift. Most of our
sample is in the lower part of this range and consistent with that
of the Milky Way thick disk.  The ratio of axes for UDF disks is
$\sim3\pm1.5$, comparable to the ratio for modern thick disks.

The disks in the UDF are composed of giant clumps closely centered
on the midplane. These clumps do not appear to be a swarm of
satellite remnants in various stage of accretion, but appear to
have formed in the disks as a result of gravitational
instabilities. Their sizes are only $\sim20$\% less than the scale
heights, which suggests the disks built up over time by the
dissolution of clumps.

Modern thick disks may not be as directly related to the UDF disks
as these observations suggest.  If the entire disk grows over
time, then it is difficult to see how the thick disk can remain
old and still preserve about the same ratio of axes during growth.
In addition, if thin disks form inside the thick disks observed
here, as a later or continuing phase of gas accretion, for
example, then the UDF thick disks should shrink down to have a
size today that is comparable to the old thin disk; i.e., they
would not become the thick disks of modern galaxies.

The observation of clumpy but straight thick disks in chain and
spiral galaxies from redshift $z\sim0.5$ to 5 seems inconsistent
with the prevailing model in which thick disks form during the
violent impact heating of thin disks and by satellite debris from
this mixing. The young disks that we observe appear more
quiescent, and they appear to be thick because they have low mass
column densities (comparable to the mass column densities of
modern thick disks, but without the underlying thin disks). If
there is a phase of impact stirring to make modern thick disks out
of thin disks, then it would have to come after the UDF epoch, or
it would have to involve UDF morphologies (e.g., irregular
collision remnants) that are not included in our spiral and chain
samples.

The possibility of a selection effect like this is important to
acknowledge. The whole notion of chain galaxies as edge-on clumpy
disks could be wrong, for example, even though the supposed
face-on counterparts are observed and in the right proportion for
inclination effects. Still, the galaxy sample used for study in
this paper was chosen according to morphology with the thinnest
members of each class used to represent the edge-on cases. If
these thin galaxies are only the more relaxed members of a larger
set of galaxies whose thicker vertical extents are determined by
violent collisions and not face-on inclinations, then our sample
would be biased.  Such a bias would affect our conclusion that UDF
thick disks formed by the dissolution of giant clusters which, in
turn, formed by in-plane gravitational instabilities. Many of the
giant clusters could be satellites in this case, forming outside
of the disk, and then the whole process of thick disk formation
could be more irregular and more violent than our sample
indicates.

Based on observations with the NASA/ESA Hubble Space Telescope,
obtained at the Space Telescope Science Institute, which is
operated by the Association of Universities for Research in
Astronomy, Inc., under NASA contract NAS 5-26555. D.M.E. thanks
Vassar College for a publication grant from the Research
Committee. We are grateful to the referee for useful comments.


\clearpage

\newpage
\newpage
\newpage
\newpage
\newpage
\newpage
\newpage
\newpage
\newpage
\newpage
\newpage
\newpage
\newpage
\newpage
\newpage
\newpage
\newpage

\end{document}